\documentstyle[12pt,axodraw,epsf]{article} 
\catcode`@=11
\newcount\@tempcntc
\def\@citex[#1]#2{\if@filesw\immediate\write\@auxout{\string\citation{#2}}\fi
  \@tempcnta\z@\@tempcntb\m@ne\def\@citea{}\@cite{\@for\@citeb:=#2\do
    {\@ifundefined
       {b@\@citeb}{\@citeo\@tempcntb\m@ne\@citea\def\@citea{,}{\bf ?}\@warning
       {Citation `\@citeb' on page \thepage \space undefined}}%
    {\setbox\z@\hbox{\global\@tempcntc0\csname b@\@citeb\endcsname\relax}%
     \ifnum\@tempcntc=\z@ \@citeo\@tempcntb\m@ne
       \@citea\def\@citea{,}\hbox{\csname b@\@citeb\endcsname}%
     \else
      \advance\@tempcntb\@ne
      \ifnum\@tempcntb=\@tempcntc
      \else\advance\@tempcntb\m@ne\@citeo
      \@tempcnta\@tempcntc\@tempcntb\@tempcntc\fi\fi}}\@citeo}{#1}}
\def\@citeo{\ifnum\@tempcnta>\@tempcntb\else\@citea\def\@citea{,}%
  \ifnum\@tempcnta=\@tempcntb\the\@tempcnta\else
   {\advance\@tempcnta\@ne\ifnum\@tempcnta=\@tempcntb \else \def\@citea{--}\fi
    \advance\@tempcnta\m@ne\the\@tempcnta\@citea\the\@tempcntb}\fi\fi}
\catcode`@=12
\begin{document}

\begin{flushright}
CERN-TH/99-167\\
THES-TP/99-07\\
hep-ph/9906265\\
June 1999
\end{flushright}

\begin{center}
{\Large {\bf Leptogenesis in Theories with Large Extra Dimensions}}\\[1.5cm]
{\large Apostolos Pilaftsis}\\[0.35cm]
{\em Theory Division, CERN, CH-1211 Geneva 23, Switzerland}\\
{\em and}\\
{\em Department of Theoretical Physics, University of Thessaloniki,}\\
{\em GR 54006 Thessaloniki, Greece}
\end{center}
\vskip1.4cm  \centerline{\bf   ABSTRACT}
We    study the  scenario of   baryogenesis   through leptogenesis  in
higher-dimensional theories, in which the  scale of quantum gravity is
many  orders  of magnitude smaller  than the  usual  Planck mass.  The
minimal realization of these  theories includes an isosinglet neutrino
which feels the presence of  large compact dimensions, whereas all the
SM particles are localized on  a $(1+3)$-dimensional subspace.  In the
formulation  of  minimal   leptogenesis   models, we   pay  particular
attention to the existence  of Majorana spinors in higher  dimensions.
After compactification  of the extra dimensions,  we obtain a tower of
Majorana Kaluza-Klein excitations  which act as  an infinite series of
CP-violating resonators, and derive the necessary conditions for their
constructive interference.  Based on  this CP-violating mechanism,  we
find  that the decays of the  heavy Majorana excitations can produce a
leptonic asymmetry  which  is reprocessed into the   observed baryonic
asymmetry of the   Universe by means of  out-of-equilibrium  sphaleron
interactions, provided the reheat temperature is above 5 GeV.

\newpage

\setcounter{equation}{0}
\section{Introduction}

Superstring theories have been   advocated  to provide a    consistent
theoretical  framework that   could lead to  quantization  of gravity,
including its possible  unification with all  other fundamental forces
in nature. The quantum nature of gravity is expected to play a central
role at energy scales close to the Planck mass, $M_{\rm P} = 1.2\times
10^{19}$  GeV.  The  formulation of superstring  theories requires the
embedding of our  well-established  $(1+3)$-dimensional world  into  a
higher-dimensional space, in which the new  spatial dimensions must be
highly curved for both phenomenological  and theoretical reasons.   In
typical string theories,  the fundamental string scale  is generically
of   order   $M_{\rm     P}$.   However, Witten    \cite{Witten},  and
Ho$\check{{\rm r}}$ava  and Witten  \cite{HW} presented an interesting
alternative, in which the string  scale may be considerably lowered to
$\sim  10^{16}$ GeV,    thereby   enabling the  unification   of   all
interactions within the  minimal  supersymmetric model.   An analogous
scenario was subsequently discussed by  Lykken \cite{JL}, in which the
string scale  was further   lowered to  the  TeV range,\footnote{In  a
different context, Antoniadis \cite{IA} had made an earlier suggestion
of a low compactification scale of order TeV in string theories.}  but
the fundamental Planck scale was kept intact to $M_{\rm P}$.

Recently,   Arkani-Hamed,   Dimopoulos  and  Dvali    \cite{ADD}  have
considered a more radical scenario, in  which the fundamental scale of
quantum gravity, $M_F$, may be as low as few TeV, thereby proposing an
appealing  solution to the known gauge  hierarchy problem \cite{DDG0}. 
The  observed  weakness of  gravity  may  then be  attributed  to  the
presence  of  a  number $\delta$  of  large extra  spatial dimensions,
within which   only gravity can propagate  and,  most probably, fields
that are singlets under  the Standard Model  (SM) gauge group, such as
isosinglet     neutrinos    \cite{ADDM,DDG}.   This  higher     $[1  +
(3+\delta)]$-dimensional space is usually  termed bulk.  On  the other
hand, all  the   ordinary  SM  particles  live  in  the   conventional
$(1+3)$-dimensional Minkowski subspace, which is called wall.  In such
a theoretical framework, the ordinary Planck  mass $M_{\rm P}$ must be
viewed as  an effective parameter, which is   related to the genuinely
fundamental scale $M_F$ through a kind of generalized Gauss law
\begin{equation}
  \label{Gauss}
M_{\rm P}\ \approx\ M_F\, (R\, M_F)^{\delta/2}\, ,      
\end{equation}
where  we  have     assumed, for   simplicity,   that the   additional
$\delta$-dimensional volume has the configuration of a torus, with all
of      its    radii      being    equal.        Many    astrophysical
\cite{ADD1,astro,ADKM,BD,CP,HS,RSun,DS}       and     phenomenological
\cite{pheno} analyses have  already appeared in  the recent literature
for such low string-scale theories.

As   has been  mentioned  already,  it  is conceivable  to assume that
isosinglet neutrinos exist  in  addition to  gravitons, and  that also
feel the presence of large extra  space dimensions.  In particular, we
wish to      study novel scenarios,   in    which  the   existence  of
higher-dimensional   singlet  fields   may account    for the observed
baryonic  asymmetry    of   the Universe    (BAU) by   means   of  the
Fukugita--Yanagida mechanism of  leptogenesis \cite{FY}.  According to
this  mechanism, an  excess  of the   lepton  number  ($L$)  is  first
generated by out-of-equilibrium $L$-violating decays of heavy Majorana
neutrinos, which is  then converted  into an  asymmetry of the  baryon
number   ($B$)   through   $(B+L)$-violating  sphaleron   interactions
\cite{KRS}.   Such  an  $L$-to-$B$ conversion    of asymmetries  stays
unsuppressed, as long as  the heavy Majorana-neutrino masses lie above
the  critical temperature  $T_c$  of the  electroweak phase transition
where sphalerons  are supposed to  be in  thermal equilibrium.  Such a
scenario  of the BAU  generation  is often called baryogenesis through
leptogenesis.

The presence   of  large  extra dimensions   introduces a    number of
alternatives  for leptogenesis which may  even have no analogue in the
conventional 4-dimensional theories.   We shall focus our attention on
minimal realizations  of higher-dimensional leptogenesis models, which
lead, after compactification of the   extra dimensions, to   scenarios
that admit renormalization  assuming a finite number of  Kaluza--Klein
(KK) excitations. Such   models of leptogenesis  are therefore endowed
with enhanced  predictive  power.  For definiteness, we  will consider
minimal  4-dimensional extensions of the  SM, augmented by one singlet
Dirac neutrino, which propagates   in the bulk.  Parenthetically,   we
should notice  that massive  Majorana  neutrinos are not   defined for
spaces with any space-time dimensions  but only  for those with  $2,3$
and $4\, {\rm  mod}\,  8$  dimensions \cite{JS,PvN}.    For  instance,
unlike in  4 dimensions, true Majorana  spinors cannot be defined in 5
dimensions.  This topic will be discussed in detail in Section 2.

After  compactification of the extra   dimensions, the kinetic term of
the bulk neutrino  gives  rise to an   infinite series  of massive  KK
excitations,    with equally spaced   Dirac    masses, i.e.\ the  mass
difference between two  neighbouring KK states is of  order $1/R$.  In
order to make the mechanism of leptogenesis work, it is necessary that
the model  under consideration violate both  the lepton number $L$ and
the product of symmetries of charge conjugation (C) and (parity) space
reflection (P), also known as CP symmetry. The violation of $L$ can be
introduced into the theory   by simultaneously coupling  the different
spinorial   states  of the higher-dimensional    Dirac neutrino to the
lepton  doublets of the SM and  to their C-conjugate counterparts.  As
we will see  in Section 3,  however, this is   not sufficient for  the
theory   to be CP-violating.    CP non-conservation  can  be minimally
realized  in  two different  ways:  one   has  to either  (i)  include
additional higher-dimensional fermionic  bilinears or  (ii) extend the
Higgs  sector  of  the SM.  Obviously,   one  may also  consider  more
involved models based on combinations of these  two minimal scenarios. 
The first scenario  may be regarded  as a higher-dimensional extension
of the ordinary  leptogenesis model \cite{FY}.   Of most  interest is,
however,  the second  alternative,    which  has  no analogue   in   4
dimensions, as it does not require the inclusion of any explicit heavy
Majorana  or   isosinglet    mass  scale in     the   Lagrangian.  The
characteristic  feature of these extensions is  that each of the Dirac
KK  neutrino states   splits   into two  nearly    degenerate Majorana
neutrinos either at the  tree level in   the first scenario or at  one
loop in the second one.

There  are generically two  distinct mechanisms  that  give rise to CP
non-conservation  in the decays of   heavy Majorana KK states.  In  the
first mechanism,  CP violation is induced  by the  interference of the
tree-level decay  graph with the  absorptive part of a one-loop vertex
diagram \cite{FY,MAL};    we call  the  latter  $\varepsilon'$-type CP
violation in connection with  the established terminology of the  kaon
system.  In the second  mechanism, which we call $\varepsilon$-type CP
violation, the tree-level  diagram interferes with the absorptive part
of   the one-loop self-energy  transition  between  two heavy Majorana
neutrinos  \cite{LS,APRD,CPres},   i.e.\  between  heavy   Majorana KK
states.  If the mass difference of two heavy Majorana states is of the
order   of      their  respective      widths,  the  description    of
$\varepsilon$-type  CP       violation  becomes        more     subtle
field-theoretically   \cite{APRD}.     In    this  case,  finite-order
perturbation theory no longer  applies, and one is therefore compelled
to resort   to a resummation   approach, which consistently  takes the
instability  of the mixed  heavy  Majorana states  into account.  This
issue has extensively been discussed in \cite{APRD}.

Furthermore, it was  shown \cite{APRD} that  the $\varepsilon$-type CP
violation   induced by  the  mixing   of  two nearly  degenerate heavy
Majorana states can be  resonantly enhanced up to  the order of unity. 
As we shall discuss in more detail in Section 4, an analogous dynamics
exhibits the system of  the Majorana KK excitations.  In fact, each KK
pair  of the  two  nearly  degenerate Majorana  states  behaves  as an
individual CP-violating resonator.  In this way, we shall characterize
a   two-level system  that    satisfies  the resonant  conditions   of
order-unity CP violation.  We  find that the  spacing in mass for  two
adjacent  KK pairs   of  Majorana  states   governs  the dynamics  for
constructive or destructive interference of the  complete tower of the
CP-violating  resonators. Owing to   cancellations among the different
CP-violating vertex contributions,  we can explicitly demonstrate that
$\varepsilon'$-type CP violation is vanishingly small.

A crucial requirement for successful baryogenesis through leptogenesis
is  that the temperature of reheating  $T_r$ \cite{KT} due to the late
decays of gravitons into photons be not much smaller than the critical
temperature    $T_c$,   namely   the  temperature   above    which the
$(B+L)$-violating sphalerons are in   thermal equilibrium.  If  $T_r <
T_c$,  sphalerons are out  of equilibrium,  and  the conversion of the
generated   leptonic asymmetry  into   the baryon   asymmetry  becomes
exponentially suppressed.  In particular, it has been argued \cite{BD}
that it may be difficult to obtain a reheat temperature above $T_c$ in
theories  with  a low  scale  of  quantum  gravity  $M_F$,  such  that
sphalerons can effectively reprocess an excess in $L$ into $B$.  For 2
extra large dimensions, the   authors \cite{ADD1,BD} derive   the mass
limit $M_F \stackrel{>}{{}_\sim}   100$ TeV, assuming that the  reheat
temperature is larger than  few MeV, so  as to ensure  that primordial
nucleosynthesis proceeds as usual.  This bound  is also in qualitative
agreement with recent constraints derived from considerations of rapid
supernovae  cooling due  to   graviton emission \cite{CP}  and  of the
cosmic diffuse gamma radiation \cite{HS}.

Nevertheless, several possibilities have  already been reported in the
literature  that one might  think  of  to avoid possible  difficulties
associated with a low $T_r$.  For example, one  could imagine that the
bulk singlet neutrino only resides in a subspace of a multidimensional
space spanned by a number $\delta = 6$ of extra dimensions and higher,
in which gravity  propagates \cite{ADDM}.  This  could lead  to rather
suppressed production rates of   gravitons, thus allowing much  larger
reheat  temperatures \cite{ADD1,BD}.     Another  way of   solving the
problem of a low $T_{r}$ is to assume  that the compactification radii
\cite{RSun,DS} of gravity   are not all   equal  but possess  a  large
hierarchy.     Such a possibility  would completely   change the usual
cosmological picture of the   previous analyses.  In this  context, it
has  been further advocated  that  gravitons might  decay faster  on a
hidden wall than the observable wall we live  on \cite{ADD1} or even a
novel type of rapid asymmetric inflation could take place \cite{ADKM}.
Because of  the variety of the solutions  suggested in the literature,
we shall  not  put forward in  our  analysis a  specific  mechanism of
increasing the reheat  temperature close to  $T_c$.   Instead, we will
simply assume that $T_r$ is a free parameter,  and place a lower limit
on  it,   based on  the requirement that   the  observed amount of $B$
asymmetry  is    produced.  In particular,   we   shall  see  that the
resonantly enhanced CP asymmetries in the decays  of the KK states are
very important to overcome part of the low-$T_r$ problem.

The paper is organized as follows: Section 2 reviews the topic related
to the ability of defining true Majorana spinors in higher-dimensional
theories.  In   Section    3, we   formulate  minimal   renormalizable
higher-dimensional models that can  lead  to successful scenarios   of
leptogenesis.  In Section 4, we   derive the necessary conditions  for
order-unity CP asymmetries due to the constructive interference of the
tower of KK CP-violating resonators, and show that $\varepsilon'$-type
CP-violating  contributions are negligible.  In  Section 5, we give an
estimate   of  the  baryonic     asymmetry,   which arises   from    a
sphaleron-converted leptonic asymmetry,  and derive  a lower bound  on
$T_r$ and $M_F$  for  successful  baryogenesis.   Finally,  Section  6
presents our conclusions.

\setcounter{equation}{0}
\section{Majorana spinors in higher dimensions}

The     violation of the   lepton number    in leptogenesis models  or
supersymmetric  theories  is naturally   mediated by Majorana  fields,
e.g.\ heavy Majorana neutrinos, neutralinos etc. The KK formulation of
these theories  necessitates an analogous extension   of the notion of
the Majorana spinor  to higher dimensions  \cite{JS,PvN}.  The ability
of defining true Majorana neutrinos in any dimensions plays a key role
in the construction of higher-dimensional leptogenesis models. Here we
shall review this topic from a more practical, for our purposes, point
of view.

We shall consider $d$-dimensional theories with one time component and
$d-1$ spatial  ones.  We assume that   the Lagrangian describing these
theories is invariant under the generalized Lorentz transformations of
the SO$(1,d-1)$ group.   In such an extended $d$-dimensional Minkowski
space, the corresponding Clifford algebra reads
\begin{equation}
  \label{Cld}
\{ \gamma^{(d)}_\mu\,,\, \gamma^{(d)}_\nu \}\ =\ 
                                     2\,g^{(d)}_{\mu\nu}\, {\bf 1}\, ,
\end{equation}
where $g^{(d)}_{\mu\nu} = {\rm diag}\, (+1,-1,\dots,-1)$, for $\mu,\nu
= 0,1,\dots,d-1$, and  $\gamma^{(d)}_\mu$ are the generalized  Dirac
gamma matrices. The  construction of these matrices  to any number  of
dimensions may   be found  recursively.    Our starting  point  is the
representation of gamma matrices for $d=2$ and $d=3$, i.e.\ 
\begin{equation}
  \label{d23}
\gamma^{(2,3)}_0\ =\ \left[ \begin{array}{cc} 0 & 1 \\ 1 & 0
  \end{array} \right] \, ,\quad
\gamma^{(2,3)}_1\ =\ \left[ \begin{array}{cc} 0 & -1 \\ 1 & 0
  \end{array} \right] \, ,\quad
\gamma^{(3)}_2\ =\ \left[ \begin{array}{cc} i & 0 \\ 0 & -i
  \end{array} \right] \, .
\end{equation}
The procedure for constructing gamma  matrices to higher dimensions is
then as follows. If $d=2m$ ($m =1,2,\dots$), we may then define
\begin{eqnarray}
  \label{gd1}
\gamma^{(d)}_0\! & =&\! \left[ \begin{array}{cc} 
0 & {\bf 1}_m \\ 
{\bf 1}_m & 0 \end{array} \right] \, ,\quad
\gamma^{(d)}_k \ = \ \left[ \begin{array}{cc} 
0 & \gamma^{(d-1)}_0 \gamma^{(d-1)}_k \\ 
-\gamma^{(d-1)}_0 \gamma^{(d-1)}_k & 0 \end{array} \right] \, ;\ 
k = 1,\dots, d-2\,, \quad\\[0.5cm]
  \label{gd2}
\gamma^{(d)}_{d-1} \! &=& \! \left[ \begin{array}{cc} 
0 & \gamma^{(d-1)}_0  \\ 
-\gamma^{(d-1)}_0  & 0 \end{array} \right] \, \ {\rm and}\quad
\gamma^{(d)}_P\ =\  \left[ \begin{array}{cc} 
{\bf 1}_m & 0   \\ 
0 & -{\bf 1}_m \end{array} \right] \, ,
\end{eqnarray}
where ${\bf 1}_m$  is the unity  matrix in $m$ dimensions.  Note  that
the dimensionality of the representation  of the gamma matrices for $d
=  2m + 1$ coincides with  that of $d=2m$. The matrix $\gamma^{(d)}_P$
is   the generalization of   the   usual  $\gamma_5$ matrix in    four
dimensions,  i.e.\  $\gamma^{(d)}_P     =  c \prod_{\mu    =  0}^{d-1}
\gamma^{(d)}_\mu$, where the  constant  $c$   is defined  such    that
$\gamma^{(d)\,2}_P =   {\bf   1}$.    The   matrix    $\gamma^{(d)}_P$
anticommutes   with all $\gamma^{(d)}_\mu$ for   $d =  2m$, whereas it
commutes  with all $\gamma^{(d)}_\mu$ for   $d  = 2m+1$, i.e.\ it   is
proportional  to the  unity matrix.  If  we  know the representation of
gamma  matrices for $d=2m$,  we can easily construct the corresponding
one for $d=2m+1$, just by including
\begin{equation}
  \label{gd3}
\gamma^{(d+1)}_d\ =\ i\,\gamma^{(d)}_P\, .
\end{equation}
In  fact, Eqs.\ (\ref{gd1})--(\ref{gd3})  are  sufficient to construct
all $\gamma^{(d)}_\mu$ in any  number $d$ of dimensions, starting from
the known expressions (\ref{d23})  for $d=2,3$. In addition, we should
notice that the  adopted representations of  $\gamma^{(d)}_\mu$ are of
the Weyl type, having the properties
\begin{equation}
  \label{gprop}
\gamma^{(d)}_0\ =\ \gamma^{(d)\,\dagger}_0\,,\qquad
\gamma^{(d)}_k\ =\ -\,\gamma^{(d)\,\dagger}_k\, ;\quad
k = 1,\dots, d-1\, .
\end{equation}
Finally, a useful property of the above  construction is the fact that
$\gamma^{(d)}_\mu$ are  self-adjoint  under  the known  bar operation,
i.e.
\begin{equation}
  \label{gbar}
\overline{\gamma}^{(d)}_\mu \ \equiv\ \gamma^{(d)}_0\,
\gamma^{(d)\,\dagger}_\mu\, \gamma^{(d)}_0\ =\ \gamma^{(d)}_\mu\, ,
\end{equation}
and $\overline{i\gamma}^{(d)}_P = i \gamma^{(d)}_P$.

Let us now define  by $\psi (x)$ a massive  fermionic free field  in a
multidimensional Minkowski space,    which satisfies  the free   Dirac
equation  of motion, i.e.\ $(  i \gamma^\mu \partial_\mu  - m ) \psi =
0$. Here and henceforth, we shall  drop the superscript `$(d)$' on the
gamma matrices to simplify notation.  The Lorentz adjoint of $\psi$ is
then given by  $\bar{\psi} = \psi^\dagger  \gamma_0$, while invariance
of    the Dirac equation    under  generalized  Lorentz tranformations
requires
\begin{equation}
  \label{Lg0}
\bar{S}\ =\ \gamma_0 S^\dagger \gamma_0\ =\ S^{-1}\, ,
\end{equation}
where 
\begin{equation}
  \label{Swn}
S \ =\ \exp \Big( -\frac{i}{4}\, \sum_n\, 
                    \omega_n \sigma_{\mu\nu} I^{\mu\nu}_n\, \Big)\, ,
\end{equation}
with $\sigma_{\mu\nu} = \frac{i}{2}\, [  \gamma_\mu ,\gamma_\nu ]$, is
the $d$-dimensional  spinorial representation of  an arbitrary Lorentz
rotation with angles $\omega_n$, and $I^{\mu\nu}_n$ are the generators
of SO$(1,d-1)$. It is easy to see that  Eq.\ (\ref{Lg0}) is equivalent
to $\gamma_0 \sigma^\dagger_{\mu\nu}\gamma_0  = \sigma_{\mu\nu}$.  The
last equality is true by virtue of Eq.\ (\ref{gbar}).

To define   charge-conjugate fermionic fields   in theories  with many
dimensions, we proceed as follows.  We start  with the classical Dirac
equation  by including  a   background  electromagnetic field  $A_\mu$
coupled to $\psi$, i.e.\ $[ i\gamma^\mu (\partial_\mu + e A_\mu) - m ]
\psi =  0$,  and then seek  for   a solution of  the respective  Dirac
equation for the antiparticle field, denoted  as $\psi^C$, which is of
the form $[ i \gamma^\mu  (\partial_\mu - e A_\mu) -  m ] \psi^C = 0$. 
In case  $\psi$ is  a  neutral  field,  e.g.\ a  neutrino,  one should
initially assume that $e\not = 0$ and then take the  limit $e\to 0$ at
the very end of the consideration.  In this way, we find that $\psi^C$
may be determined in terms of $\psi$ as follows:
\begin{equation}
  \label{psiC}
\psi^C\ =\ C\, \bar{\psi}^T\ =\ C\, \gamma_0\, \psi^*\, ,
\end{equation}
where $C$   is  the charge-conjugation   operator  that satisfies  the
property
\begin{equation}
  \label{Coper1}
C^{-1}\, \gamma_\mu\, C\ =\ -\, \gamma^T_\mu\, ,
\end{equation}
for massive fermionic fields. For massless fermions, we may also allow
the equality
\begin{equation}
  \label{Coper2}
C^{-1}\, \gamma_\mu\, C\ =\ \gamma^T_\mu\, .
\end{equation}
Furthermore, consistency of charge conjugation with Lorentz invariance
implies that
\begin{equation}
  \label{CLor}
C^{-1}\, S\, C\ =\ (S^{-1})^T\, ,
\end{equation}
or equivalently that $C^{-1}\sigma_{\mu\nu}C =    -\sigma^T_{\mu\nu}$,
which holds true because of Eq.\ (\ref{Coper1}) or (\ref{Coper2}).  At
this point, we should remark that the transformations
\begin{equation}
   \label{Umatr}
\gamma'_\mu\ =\ U\,\gamma_\mu\, U^{-1}\,,\qquad
C'\ =\ U\,C\,U^T
\end{equation}
preserve  all the relations of  gamma  matrices given above, including
Eqs.\ (\ref{Lg0}) and (\ref{CLor}).

The necessary and sufficient condition for the existence of a Majorana
spinor in any number of dimensions reads
\begin{equation}
  \label{Maj}
\psi\ =\ \psi^C\, ,
\end{equation}
which amounts to 
\begin{equation}
  \label{CMaj}
C\,\gamma_0^T\, (C\gamma_0^T)^*\ =\ 1\, .
\end{equation}
This last equality may be rewritten as
\begin{equation}
  \label{Cg0}
C^{-1}\, \gamma_0\, C\ =\ (C^* C )\, \gamma_0^T\, .
\end{equation}
Consequently, massive (massless)  Majorana  spinors in $d$  dimensions
are admitted, if both the construction of a $C$ matrix satisfying Eq.\ 
(\ref{Coper1}) (Eq.\ (\ref{Coper2})) and $C^*  C = -1$  ($C^*C = + 1$)
is possible. As we will see below, this is not always the case.

For this purpose,  it is important to  be  able to construct  a matrix
that obeys the identity  (\ref{Coper1}) or (\ref{Coper2}).   There are
only two candidates that could be of interest:
\begin{eqnarray}
  \label{CA}
C_A \! &=&\! \prod_i^p\, \gamma_i\, ,\quad {\rm with}\quad 
\gamma_i = -\gamma^T_i = -\gamma^\dagger_i\, ,\\
  \label{CS}
C_S \! &=&\! \prod_r^s\, \gamma_r\, ,\quad {\rm with}\quad 
\gamma_r = \gamma^T_r\, ,\ \gamma_0 = \gamma^\dagger_0\, ,\
\gamma_r = -\gamma^\dagger_r\ (r\neq 0)\, .
\end{eqnarray}
Specifically, $C_A$ ($C_S$) is formed by  the product of all $p$ ($s$)
in number gamma matrices   that are purely antisymmetric (symmetric).  
Employing the identity: $\gamma_\mu  \gamma^\dagger_\mu = {\bf 1}$, we
can easily  find the following relations  for the  two $C$-conjugation
matrices:
\begin{eqnarray}
  \label{CAp}
C^{-1}_A \!\!\!&=&\!\!\! C^\dagger_A = (-1)^p\, \varepsilon (p) C_A\,,\quad
C^T_A = (C^\dagger_A)^* = (-1)^p\, \varepsilon (p) C^*_A\,,\quad
C_A = C^*_A\,,\\
  \label{CSp}
C^{-1}_S \!\!\!&=&\!\!\! C^\dagger_S = (-1)^{s-1}\, \varepsilon (s) C_S\,,\quad
C^T_S = (C^\dagger_S)^* = (-1)^{s-1}\, \varepsilon (s) C^*_S\,,\quad
C^*_S = (-1)^{s-1}\,C_S\,,\qquad
\end{eqnarray}
with $\varepsilon (z)  = (-1)^{z(z-1)/2}$.  As  advertised,  it can be
shown that the two $C$-conjugation matrices satisfy the relations
\begin{equation}
  \label{CACS}
C^{-1}_A\, \gamma_\mu\, C_A\ =\ (-1)^p\,\gamma^T_\mu\,,\qquad
C^{-1}_S\, \gamma_\mu\, C_S\ =\ (-1)^{s+1}\,\gamma^T_\mu\, .
\end{equation}
\begin{table}[t]

\begin{center}

\begin{tabular}{|r|c r||c r||c|}
\hline
$d$ & $s$  & $\varepsilon (s) = -1$  & $p$  & $\varepsilon (p) = 1$ 
                                               & Existence of massive \\
     & even & & odd &  & Majorana spinor \\
\hline\hline
 2 &  1 &  1 ~~~~&  1 &  1 ~~~~& yes \\
 3 &  2 & --1 ~~~~&  1 &  1 ~~~~& yes \\
 4 &  2 & --1 ~~~~&  2 & --1 ~~~~& yes \\
 5 &  3 & --1 ~~~~&  2 & --1 ~~~~& no  \\
 6 &  3 & --1 ~~~~&  3 & --1 ~~~~& no  \\
 7 &  4 &  1 ~~~~&  3 & --1 ~~~~& no  \\
 8 &  4 &  1 ~~~~&  4 &  1 ~~~~& no  \\
 9 &  5 &  1 ~~~~&  4 &  1 ~~~~& no  \\
10 &  5 &  1 ~~~~&  5 &  1 ~~~~& yes \\
11 &  6 & --1 ~~~~&  5 &  1 ~~~~& yes \\
12 &  6 & --1 ~~~~&  6 & --1 ~~~~& yes \\
13 &  7 & --1 ~~~~&  6 & --1 ~~~~& no  \\
\hline
\end{tabular}
\end{center}

\caption{ Existence of massive Majorana spinors in $d$ 
dimensions.}\label{Tab1}

\end{table}

On the  other hand, the  Majorana condition  given by Eq.\ (\ref{Cg0})
may now be translated into
\begin{equation}
  \label{CACSM}
C^{-1}_A\, \gamma_0\, C_A\ =\ (-1)^p\,\varepsilon (p)\, \gamma^T_0\,,\qquad
C^{-1}_S\, \gamma_0\, C_S\ =\ \varepsilon (s)\, \gamma^T_0\, .
\end{equation}
As a consequence,  the existence of a massive  Majorana spinor  in any
number of dimensions is ensured, if
\begin{equation}
  \label{restr1}
\varepsilon (p) = 1\quad {\rm and} \quad p\ {\rm is\ odd}\, ,
\end{equation}
or if 
\begin{equation}
  \label{restr2}
\varepsilon (s) = -1\quad {\rm and} \quad s\ {\rm is\ even}\, .
\end{equation}
Based on Eqs.\ (\ref{restr1})    and (\ref{restr2}), we  can  generate
Table \ref{Tab1}.  As can be seen from  this table, true {\em massive}
Majorana neutrinos exist   only in  $2,3$   and $4\, {\rm  mod}\,   8$
dimensions \cite{JS}.
\begin{table}[t]

\begin{center}

\begin{tabular}{|r| r|| r c r||c|}
\hline
$d$ & $\varepsilon (p) = 1$  & $\varepsilon (s) $ & $=$ & $(-1)^{s+1}$ 
                                          & Existence of massless \\
     &   & &                   &          & Majorana spinor \\
\hline\hline
 5 & --1 & --1 & &  1 ~~~~& no  \\
 6 & --1 & --1 & &  1 ~~~~& no  \\
 7 & --1 &  1 & & --1 ~~~~& no  \\
 8 &  1 &  1 & & --1 ~~~~& yes  \\
 9 &  1 &  1 & &  1 ~~~~& yes \\
\hline
\end{tabular}
\end{center}

\caption{ Existence of massless Majorana-Weyl spinors in $d$ 
dimensions.}\label{Tab2}

\end{table}

If we also allow the possibility of massless Majorana-Weyl spinors, we
only need to impose the restriction:
\begin{equation}
  \label{restr3}
\varepsilon (p) = +1\quad {\rm or}\quad \varepsilon (s) = (-1)^{s+1}\,.
\end{equation}
Generated in this way, Table \ref{Tab2} shows  that in addition to the
result found in the massive case, the  definition of massless Majorana
fields  can  be  extended to 8    and $9\, {\rm  mod}\,  8$ dimensions
\cite{PvN}.  For example, our  analysis explicitly demonstrates  that,
as opposed to theories  with  4  dimensions, true  Majorana  neutrinos
cannot be defined  in those with $5,\  6$, and 7 dimensions.  In fact,
in the latter theories, $C$ loses its very meaning  of being a genuine
charge-conjugation matrix.  We shall pay   special attention to   this
issue in the next  section, while formulating different minimal models
of leptogenesis.

\setcounter{equation}{0}
\section{Higher-dimensional models of leptogenesis}

If    the SM  contains a  singlet   neutrino  that  feels large  extra
dimensions,   this  additional   volume factor   of   the  new spatial
dimensions introduces a  new   possibility to naturally   suppress the
Higgs Yukawa coupling  to neutrinos \cite{ADDM,DDG}. After spontaneous
symmetry    breaking (SSB) of  the  SM  Higgs potential, the resulting
neutrino masses  may naturally be of the  order of $10^{-2}$ eV, which
turns out to be  in the  right  ballpark for explaining the  solar and
atmospheric neutrino data  \cite{DS}.  Here we shall formulate minimal
models of  leptogenesis  which,  after compactification of   the extra
dimensions,  give rise to  theories containing 4-dimensional operators
only, and  are therefore renormalizable for  finitely many KK  states. 
Even though  the  number of  KK excitations  is formally  infinite, on
theoretical  grounds, however,  one   expects   the  presence  of   an
ultra-violet  (UV) cutoff close to the  string scale  where gravity is
supposed to set in.  The issue of renormalization will become clearer,
when we describe the leptogenesis models.

For  simplicity,  we  shall   consider a    5-dimensional  model.  The
generalization  of  the  results  to     higher dimensions   is   then
straightforward.  Following   \cite{ADDM,DDG},   we  assume  that  all
particles  with  non-zero SM charges   live  in a  subspace of $(1+3)$
dimensions.    Also,  we   introduce one   Dirac  isosinglet  neutrino
$N(x_\mu,y)$ that   propagates in the bulk  of  all 5  dimensions.  We
denote by $x_\mu  = (x_0, x_1, x_2, x_3)$  the one time and the  three
spatial coordinates of  our observable world and  by $y\equiv x_4$ the
new spatial dimension.  The $y$-coordinate is  to be compactified on a
circle  of  radius  $R$ by  applying the  periodic  identification: $y
\equiv y  + 2\pi R$.    Specifically, the minimal  field  content of a
one-generation model of leptogenesis is
\begin{equation}
  \label{content}
L(x)\ =\ \left( \begin{array}{c} \nu_L (x) \\ l_L (x) \end{array}
\right) ,\qquad
l_R (x),\qquad N(x,y)\ =\ \left( \begin{array}{c} \xi (x,y) \\ 
\bar{\eta} (x,y) \end{array} \right)\, ,
\end{equation}
where   $\nu_L$, $l_L$ and $l_R$  are  4-dimensional Weyl spinors, and
$\xi$ and $\eta$ are two-component spinors in 5 dimensions.  Depending
on the model,  we shall also  assume that $\xi$ ($\eta$)  is symmetric
(antisymmetric) under  a $y$ reflection:  $\xi (x,y) = \xi (x,-y)$ and
$\eta (x,y)  = -\eta (x,-y)$.\footnote{With  the imposition of  such a
  symmetry  which might be   justified within the  context  of a $Z_2$
  orbifold compactification  \cite{DDG}, a twofold mass degeneracy may
  be avoided in the  spectrum of the KK   states for the  leptogenesis
  models under study.}  Following the procedure outlined in Section 2,
the gamma matrices in 5 dimensions may be represented by
\begin{equation}
  \label{gam5}
\gamma_\mu\ =\ \left( \begin{array}{cc} 0 & \bar{\sigma}_\mu \\
                \sigma_\mu & 0 \end{array} \right) ,\quad {\rm and}\quad
\gamma_4\ =\ \left( \begin{array}{cc} i{\bf 1}_2 & 0 \\
                                      0 & -i{\bf 1}_2 \end{array}\right),
\end{equation}
where $\sigma^\mu = ({\bf  1}_2, \vec{\sigma} )$ and $\bar{\sigma}^\mu
= ({\bf 1}_2, -\vec{\sigma}  )$, with $\vec{\sigma}_{1,2,3}$ the usual
Pauli matrices.

As we have mentioned in the introduction, there are two representative
minimal scenarios of leptogenesis:
\begin{itemize}
  
\item[ (i)]  The first scenario  may be viewed as a higher-dimensional
  generalization of the usual  leptogenesis model of Ref.\  \cite{FY},
  in which  the   Lorentz- and gauge-  invariant  fermionic  bilinears
  $\bar{N}  N$ and $N^T  C^{(5)-1} N$ are  included.   As we will see,
  however, if a $Z_2$ discrete  symmetry  is imposed on $N(x,y)$,  the
  former bilinear mass term $\bar{N}   N$ does not contribute to   the
  effective action.  According to Eqs.\ (\ref{CA}) and (\ref{CS}), the
  matrix  $C^{(5)}$ satisfies  Eqs.\ (\ref{Coper2}) and  (\ref{CLor}),
  but not Eq.\ (\ref{Coper1}), which  defines the true $C$-conjugation
  matrix for a massive  Dirac field.  Despite its  close analogy  to 4
  dimensions, the  operator  $N^T C^{(5)-1}  N$  does not  represent a
  genuine bare Majorana mass  in 5 dimensions.  Nevertheless, after KK
  compactification, the effective Lagrangian of this scenario displays
  a dynamics  rather analogous to  the known scenario  of leptogenesis
  due to Fukugita and Yanagida \cite{FY}.
  
\item[(ii)] The second  scenario of leptogenesis requires, in addition
  to the bulk  Dirac singlet field $N(x,y)$,  that the Higgs sector of
  the  SM be extended  by two  more Higgs  doublets.  The  first Higgs
  doublet, denoted as $\Phi_1$,  couples to the lepton isodoublet $L$,
  the second  Higgs  doublet $\Phi_2$ couples  to its charge-conjugate
  counterpart, $C\bar{L}^T$,  while  the  last  one  $\Phi_3$  has  no
  coupling to  matter.  Thus, the  extended Higgs potential  admits CP
  non-conservation, which originates  from the bilinear mixing  of the
  three Higgs doublets.   In  fact, in  this model, both  the Majorana
  masses of the KK excitations and CP violation are generated via loop
  effects.  Most interestingly, as we will detail below, this scenario
  of leptogenesis has no analogue in 4 dimensions.

\end{itemize}

Of course, one may consider  more involved models of leptogenesis that
are  based  on combinations of   the   basic scenarios  (i) and  (ii),
including their  possible supersymmetric extensions.  Therefore, it is
very  instructive to analyze in  more  detail these two representative
models of leptogenesis, as  well  as a  hybrid scenario that  includes
both the extensions mentioned above, i.e.\ fermionic bilinears and two
additional Higgs doublets.

\subsection{The leptogenesis model with fermionic bilinears}

In this scenario,  the SM is augmented  by a higher-dimensional  Dirac
singlet neutrino $N(x,y)$, while the SM particles are considered to be
confined to a 4-dimensional hypersurface,   which describes our  world
and is   often  termed as   a 3-brane.\footnote{In  a  field-theoretic
  context,  Rubakov     and  Shaposhnikov \cite{RS}    presented   the
  possibility  of dynamically localizing  4-dimensional  fermions on a
  solitonic brane embedded in a higher-dimensional space, by employing
  the  index theorem in a  solitonic background  \cite{JRW}.}  In this
picture, the bulk Dirac neutrino field $N(x,y)$ intersects the 3-brane
at a position  $y  = a$, which naturally   gives rise to small  Yukawa
couplings suppressed  by the  volume of the  extra  dimensions.   This
suppression mechanism is very much like  the one that gravity owes its
weakness at  long distances  in theories with  a low  scale of quantum
gravity  \cite{ADD}.  The most general   effective  Lagrangian of  the
scenario under discussion is given by
\begin{eqnarray}
  \label{Leff1}
{\cal L}_{\rm eff}  & =& \int\limits_0^{2\pi R}\!\! dy\
 \Big\{\, \bar{N} \Big( i\gamma^\mu \partial_\mu\, +\, 
 i\gamma_4 \partial_y \Big) N\ -\ m \bar{N} N\ -\ 
\frac{1}{2}\,\Big( M N^T C^{(5)-1} N\ +\
 {\rm H.c.} \Big) \nonumber\\
&&+\,\delta (y-a)\, \Big[\, \bar{h}_1\, L\tilde{\Phi} \xi\, +\,
\bar{h}_2\, L \tilde{\Phi} \eta\ +\ {\rm H.c.}\,\Big]\ +\ 
\delta (y-a)\, {\cal L}_{\rm SM}\, \Big\}\, ,
\end{eqnarray}
where ${\cal L}_{\rm SM}$ denotes the SM Lagrangian and 
\begin{equation}
  \label{C5}
C^{(5)}\ =\ - \gamma_1 \gamma_3\ =\ \gamma_0  \gamma_2 \gamma_4\ =\ \left[
\begin{array}{cc} -i\sigma_2 & 0 \\
  0 & -i\sigma_2 \end{array} \right]\, .
\end{equation}
In Eq.\  (\ref{Leff1}),  $\tilde{\Phi}  =  i\sigma_2  \Phi^*$ is   the
hypercharge-conjugate of the SM Higgs  doublet,  and $\xi$ and  $\eta$
are    higher-dimensional  two-component  spinors   defined   in  Eq.\
(\ref{content}).   Note   that    $\bar{h}_1$   and  $\bar{h}_2$   are
dimensionful   kinematic parameters,  which    may be related   to the
dimensionless Yukawa couplings $h_1$ and $h_2$ through
\begin{equation}
  \label{h5h4}
\bar{h}_{1,2}\ =\ \frac{h_{1,2}}{(M_F)^{\delta /2}}\ ,
\end{equation}
with $\delta =1$. Here, one must remark  that the fundamental scale of
quantum gravity $M_F$ occurs naturally in  Eq.\ (\ref{h5h4}), as it is
the only  available energy  scale    of the effective   Lagrangian  to
normalize these higher-dimensional Yukawa couplings.

Given that $N(x,y)$ is a periodic function of $y$, with a period $2\pi
R$ and its  two-component  spinorial  modes being constrained  by  the
aforementioned $Z_2$  discrete symmetry \cite{DDG},  we may expand the
two-component spinors $\xi$ and $\eta$ in a Fourier series as follows:
\begin{eqnarray}
  \label{xi}
\xi (x,y) &=& \frac{1}{\sqrt{2\pi R}}\ \xi_0 (x)\ +\ 
\frac{1}{\sqrt{\pi R}}\ \sum_{n=1}^\infty\, \xi_n (x)\ 
                                       \cos\bigg(\,\frac{ny}{R}\,\bigg)\,,\\
  \label{eta}
\eta (x,y) & =& \frac{1}{\sqrt{\pi R}}\ \sum_{n=1}^\infty\, \eta_n (x)\ 
                                       \sin\bigg(\,\frac{ny}{R}\,\bigg)\,,
\end{eqnarray}
where the chiral spinors $\xi_n (x)$ and $\eta_n (x)$ form an infinite
tower of   KK  modes. After  integrating out  the  $y$ coordinate, the
effective Lagrangian takes on the form
\begin{eqnarray}
  \label{Leff1KK}
{\cal L}_{\rm eff} & = & {\cal L}_{\rm SM}\ +\ \bar{\xi}_0
( i\bar{\sigma}^\mu \partial_\mu) \xi_0\ 
+\ \Big(\, \bar{h}^{(0)}_1\, L\tilde{\Phi} \xi_0\ -\
\frac{1}{2}\, M\, \xi_0\xi_0\ +\ {\rm H.c.}\,\Big)\
 +\ \sum_{n=1}^\infty\, \bigg[\, \bar{\xi}_n 
( i\bar{\sigma}^\mu \partial_\mu) \xi_n\nonumber\\
&& +\, 
\bar{\eta}_n ( i\bar{\sigma}^\mu \partial_\mu) \eta_n\
+\ \frac{n}{R}\, \Big( \xi_n \eta_n\, +\, \bar{\xi}_n
\bar{\eta}_n\Big) -\ \frac{1}{2}\, M\, 
\Big( \xi_n\xi_n\, +\, \bar{\eta}_n\bar{\eta}_n\
+\ {\rm H.c.}\Big)\nonumber\\
&& +\, \sqrt{2}\, \Big(\, \bar{h}^{(n)}_1\, L\tilde{\Phi} \xi_n\ +\
\bar{h}^{(n)}_2\, L\tilde{\Phi} \eta_n\ +\ {\rm H.c.}\,\Big)\, \bigg]\, ,
\end{eqnarray}
where we have chosen the weak basis in which $M$ is positive, and
\begin{eqnarray}
  \label{h1n}
\bar{h}^{(n)}_1 &=& \frac{h_1}{(2\pi M_F R)^{\delta/2}}\ 
\cos \bigg(\,\frac{na}{R}\,\bigg)\ =\ \frac{M_F}{M_{\rm P}}\  
h_1 \cos \bigg(\,\frac{na}{R}\,\bigg)\qquad (n \ge 0)\,,\\
  \label{h2n}
\bar{h}^{(n)}_2 &=& \frac{h_2}{(2\pi M_F R)^{\delta/2}}\ 
\sin \bigg(\,\frac{na}{R}\,\bigg)\ =\ \frac{M_F}{M_{\rm P}}\  
h_2 \sin \bigg(\,\frac{na}{R}\,\bigg)\qquad \ (n \ge 1)\, . 
\end{eqnarray}
In deriving the last equalities on  the RHS's of Eqs.\ (\ref{h1n}) and
(\ref{h2n}), we  have employed the    basic  relation given in    Eq.\ 
(\ref{Gauss}).  In agreement    with  \cite{ADDM,DDG}, we   find  that
independently of  the number  $\delta$  of the  extra  dimensions, the
4-dimensional Yukawa couplings $\bar{h}^{(n)}_1$ and $\bar{h}^{(n)}_2$
are  naturally suppressed  by  an extra volume  factor $M_F/M_{\rm  P}
\stackrel{<}{{}_\sim}  10^{-10}$.  We also  observe that the mass term
$m \bar{N} N$ drops out from the effective Lagrangian,  as a result of
the $Z_2$ discrete symmetry.

In  the symmetric  (unbroken)  phase of  the  theory, the  part of the
Lagrangian describing the KK masses is given by
\begin{eqnarray}
  \label{MKK1}
-\, {\cal L}^{\rm KK}_{\rm mass} \!\!& = &\!\! \frac{1}{2}\, M\, \xi_0\xi_0\
+\ \frac{1}{2}\ \sum_{n=1}^{\infty}\,
\Big(\, \xi_n\,,\ \eta_n\, \Big)
\left( \begin{array}{cc}
M & - n/R \\ -n/R & M \end{array} \right)
\left( \begin{array}{c} \xi_n \\ \eta_n \end{array} \right)\ +\  
                             {\rm H.c.} \\
\!\!& = &\!\! \frac{1}{2}\, \mu\, \chi^{(0)}_1\chi^{(0)}_1\
+\ \frac{1}{2}\ \sum_{n=1}^{\infty}\,
\Big(\, \chi^{(n)}_1 \,,\ \chi^{(n)}_2\, \Big)
\left( \begin{array}{cc}
n/R - \mu & 0 \\ 0 & n/R + \mu  \end{array} \right)
\left( \begin{array}{c} \chi^{(n)}_1 \\ \chi^{(n)}_2 \end{array}
                                           \right)\ +\ {\rm H.c.},\nonumber
\end{eqnarray}
where $\chi^{(n)}_{1(2)} = \frac{1}{\sqrt{2}} \exp  (i\phi^n_{1(2)})\,
[\,\xi_n  +(-)\, \eta_n\,]$.  As in  Ref.\ \cite{DDG}, we have defined
$\mu = {\rm min}\,  ( |M -  \frac{k}{R}|  )$ to  be the smallest  mass
eigenvalue  for some  given value  of   $k$, and have  relabelled  the
remaining KK  mass eigenstates  $\chi^{(n)}_1$ and $\chi^{(n)}_2$ with
respect to  $k$.  Thus, after compactification,  we see how  the heavy
isosinglet mass  $M$ gets replaced by the  small  Majorana mass $\mu$,
with $\mu \le 1/R$.  Further technical details and a discussion may be
found in \cite{DDG}. After expressing the effective Lagrangian in Eq.\ 
(\ref{Leff1KK}) in the newly introduced Majorana-mass basis, we obtain
\begin{eqnarray}
  \label{LKK1}
{\cal L}_{\rm eff} \!&=&\! {\cal L}_{\rm SM}\, +\, {\cal L}^{\rm KK}_{\rm
  mass}\, +\, \bar{\chi}^{(0)}_1 ( i\bar{\sigma}^\mu \partial_\mu) 
\chi^{(0)}_1\, +\, \Big(\,h^{(0)}_1\, L\tilde{\Phi}
\chi^{(0)}_1\, +\, {\rm
  H.c.}\, \Big)\ +\, \sum_{n=1}^\infty\, \bigg[\, \bar{\chi}^{(n)}_1 
( i\bar{\sigma}^\mu \partial_\mu) \chi^{(n)}_1\nonumber\\
&&+\, \bar{\chi}^{(n)}_2 ( i\bar{\sigma}^\mu \partial_\mu) \chi^{(n)}_2\,
+\, \Big(\, h^{(n)}_1\, L \tilde{\Phi} \chi^{(n)}_1\, +\,
h^{(n)}_2\,L \tilde{\Phi} \chi^{(n)}_2\, 
+\, {\rm H.c.}\,\Big)\, \bigg]\, ,
\end{eqnarray}
where
\begin{equation}
  \label{h12}
h^{(n)}_{1(2)}\ =\ e^{i\phi^n_1}\, \bar{h}^{(n)}_1\ +(-)\ \,
e^{i\phi^n_2}\, \bar{h}^{(n)}_2\, ,
\end{equation}
and  the Yukawa couplings   $\bar{h}^{(n)}_{1,2}$  are those given  in
Eqs.\ (\ref{h1n}) and (\ref{h2n}).

It is now easy to  recognize that the  Lagrangian in Eq.\ (\ref{LKK1})
is the known  4-dimensional  model of leptogenesis, with   an infinite
number   of    pairs of     Majorana  neutrinos   $\chi^{(n)}_1$   and
$\chi^{(n)}_2$  \cite{FY}.   With   the  help of  a   method based  on
generalized CP   transformations    \cite{BBG},  we  can    derive the
sufficient and necessary condition for the theory  to be CP-invariant. 
Adapting the  result  found    in \cite{APRD}  to   the model    under
discussion, we find the condition
\begin{equation}
\label{CPinv1}
{\rm Im\, Tr}\, \Big(\, h^\dagger\, h\, \widehat{M}^{\dagger}_\chi\, 
\widehat{M}_\chi\, \widehat{M}^{\dagger}_\chi\, h^T\, h^*\, 
\widehat{M}_\chi\, \Big)\ =\ 0\, ,
\end{equation}
where  $h  = (   h^{(0)}_1,\ h^{(1)}_1,\ h^{(1)}_2,\dots,  h^{(n)}_1,\ 
h^{(n)}_2,\dots  )$ and    $\widehat{M}_\chi  =  {\rm diag}\,   (\mu,\ 
\frac{1}{R}  -\mu ,   \frac{1}{R}  + \mu,  \dots,  \frac{n}{R}  -\mu ,
\frac{n}{R} + \mu, \dots )$ are formally infinite-dimensional matrices
that contain  the Higgs Yukawa  couplings and the KK mass-eigenvalues,
respectively.  It  is a formidable  task to analytically calculate the
LHS of Eq.\  (\ref{CPinv1}).  Instead, we  notice  that if one of  the
following equalities holds true:
\begin{equation}
  \label{CPcond1}
\mu\ =\ 0\,, \qquad a\ =\ \Big(\, 0\ \, {\rm or}\ \, \pi R\, \Big)\,,
\qquad {\rm Im}\, (h_1h^*_2)^2\ =\ 0\, ,
\end{equation}
the theory is then invariant  under CP transformations.  Consequently,
CP violation requires to  have $\mu \ne 0$ and  a non-zero shifting of
the brane, $a\ne 0$, apart from  a relative CP-violating phase between
the  original  Yukawa couplings $h_1$ and   $h_2$.  Finally, we should
remark  that if   the $Z_2$  discrete symmetry  were   not imposed  on
$N(x,y)$,  the resulting Lagrangian  would predict a dangerous twofold
mass degeneracy  in the spectrum  of the would-be Majorana  KK states,
which would effectively correspond  to the $\mu  = 0$ case, and  hence
would lead to the absence of CP violation as well.

\subsection{The leptogenesis model with extended Higgs sector}

The  second scenario that  we will be discussing  does not involve the
inclusion of any heavy isosinglet mass scale.  Instead, in addition to
the higher-dimensional Dirac field $N(x,y)$  supplemented by the $Z_2$
discrete symmetry, we shall extend the  Higgs sector by two more Higgs
doublets that carry the same hypercharge as the  SM Higgs doublet.  As
we will demonstrate below, such an extension of the Higgs potential by
three  Higgs  doublets,  hereafter denoted  by $\Phi_1$,  $\Phi_2$ and
$\Phi_3$,  is dictated by the  necessity of introducing sufficient $L$
and  CP  violation into the  theory.   Specifically, this  scenario is
governed by the effective Lagrangian
\begin{eqnarray}
  \label{Leff2}
{\cal L}_{\rm eff}  & =& \int\limits_0^{2\pi R}\!\! dy\
 \Big\{\, \bar{N} \Big( i\gamma^\mu \partial_\mu\, +\, 
 i\gamma_4 \partial_y \Big) N \ +\ 
\delta (y-a)\, \Big[\, \bar{h}_1\, L\tilde{\Phi}_1 \xi\, +\,
\bar{h}_2\, L \tilde{\Phi}_2 \eta\ +\ {\rm H.c.}\,\Big]\nonumber\\
&& +\, \delta (y-a)\, \Big[\, {\cal L}'_{\rm SM} (\Phi_1 )\, +\, 
{\cal L}_V (\Phi_1,\Phi_2,\Phi_3)\,\Big]\, \Big\}\, ,
\end{eqnarray}
where $\tilde{\Phi}_i  = i\sigma_2  \Phi^*_i$ ($i=1,2,3$), and  ${\cal
L}_V  (\Phi_1,\Phi_2,\Phi_3)$   and ${\cal   L}'_{\rm  SM} (\Phi_1  )$
describe the Higgs  potential and the   residual standard part  of the
model, respectively.  Furthermore,   the model is invariant under  the
transformations:
\begin{eqnarray}
  \label{symm}
N &\to & iN\,,\qquad \Phi_1\ \to\ -i\Phi_1\,,\qquad
\Phi_2\ \to\ i\Phi_2\,,\qquad \Phi_3\ \to\ \Phi_3\,,\nonumber\\
l_R &\to & -il_R\,,\qquad u_R\ \to\ iu_R\,,\qquad d_R\ \to\ -id_R\,,
\end{eqnarray}
where $l_R$, $u_R$  and $d_R$ denote the right-handed charged-leptons,
the up- and  down-type quarks, respectively.  Obviously, only $\Phi_1$
couples to the observed SM particles, whereas $\Phi_3$ does not couple
to  matter at all. The discrete  symmetry in Eq.\ (\ref{symm}) is very
crucial,  as   it ensures  the   renormalizability of the   model; the
discrete symmetry is only broken softly by operators of dimension two:
\begin{equation}
  \label{m2ij}
 {\cal L}^{\rm soft}_V \ =\ \sum_{i<j =1}^3\ m^2_{ij}\, \Phi^\dagger_i
 \Phi_j\ +\ {\rm H.c.}\ \subset\ {\cal L}_V (\Phi_1,\Phi_2,\Phi_3)\, .
\end{equation}
Notice  that the Higgs  potential of this scenario  is very similar to
that of Weinberg's three Higgs-doublet model \cite{SW}.

One might now  naively argue that the  third Higgs doublet $\Phi_3$ is
not compelling for introducing  CP  violation into the theory,   e.g.\ 
${\rm Im}\, (\bar{h}_1 \bar{h}^*_2 m^2_{12}) \ne 0$.  However, this is
not true.  Notwithstanding $\bar{h}_1$ and $\bar{h}_2$ might initially
be  complex in the  basis in which  $m^2_{12}$ is real, one can always
rephase $L  \to e^{i\phi_l} L$ and $N   \to e^{i\phi} N$  to make both
real.  If  $\phi_{h_1}$ and $\phi_{h_2}$  denote the phases of the two
Higgs   Yukawa couplings, these  phases can  be eliminated by choosing
$\phi_l   = (\phi_{h_1} +   \phi_{h_2})/2$  and $\phi = (\phi_{h_2}  -
\phi_{h_1})/2$.  In  this  scenario,  CP violation gets   communicated
radiatively to   the  neutrino  sector through   bilinear Higgs-mixing
effects.  To be precise, CP non-conservation in the symmetric phase of
the Higgs potential ${\cal L}_V$ is manifested by the non-vanishing of
the following rephasing-invariant quantity \cite{Lavoura}:
\begin{equation}
  \label{CPbil}
{\rm Im}\, \Big(\, m^2_{12} m^2_{23} m^{*2}_{13}\, \Big)\ \ne\ 0\, .
\end{equation}
In addition, CP violation can only occur  on a shifted brane, i.e.\ $a
\ne 0$.  The latter  amounts to non-zero  values for both compactified
Higgs Yukawa couplings $\bar{h}^{(n)}_1$ and $\bar{h}^{(n)}_2$.
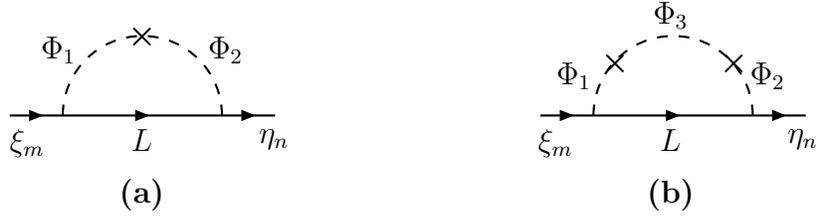
\begin{figure}
\begin{center}
\begin{picture}(300,100)(0,0)
\SetWidth{0.8}

\ArrowLine(0,30)(20,30)\ArrowLine(20,30)(80,30)\ArrowLine(80,30)(100,30)
\DashCArc(50,30)(30,0,180){4}
\Text(50,60)[]{\boldmath $\times$}
\Text(0,25)[tl]{$\xi_m$}\Text(100,25)[t]{$\eta_n$}
\Text(25,55)[r]{$\Phi_1$}\Text(75,55)[l]{$\Phi_2$}
\Text(50,25)[t]{$L$}

\Text(50,0)[]{\bf (a)}

\ArrowLine(200,30)(220,30)\ArrowLine(220,30)(280,30)\ArrowLine(280,30)(300,30)
\DashCArc(250,30)(30,0,180){4}
\Text(229,50)[]{\boldmath $\times$}\Text(274,50)[]{\boldmath $\times$}
\Text(200,25)[tl]{$\xi_m$}\Text(300,25)[t]{$\eta_n$}
\Text(220,45)[r]{$\Phi_1$}\Text(280,45)[l]{$\Phi_2$}
\Text(250,65)[b]{$\Phi_3$}
\Text(250,25)[t]{$L$}

\Text(250,0)[]{\bf (b)}

\end{picture}
\end{center}
\caption{Feynman graphs giving rise to UV-finite KK kinetic terms.}\label{f1}
\end{figure}

Proceeding as in Section 3.1,  we integrate out the compact coordinate
$y$ in Eq.\ (\ref{Leff2}) to eventually arrive at
\begin{eqnarray}
  \label{LKK2}
{\cal L}_{\rm eff} & = & {\cal L}'_{\rm SM} (\Phi_1)\ +\ 
{\cal L}_V(\Phi_1,\Phi_2,\Phi_3)\ 
+\ \bar{\xi}_0 ( i\bar{\sigma}^\mu \partial_\mu) \xi_0\ 
+\ \Big(\, \bar{h}^{(0)}_1\, L\tilde{\Phi}_1 \xi_0\ +\
{\rm H.c.}\,\Big)\nonumber\\
&& +\,
\sum_{n=1}^\infty\, \bigg[\, \bar{\xi}_n 
( i\bar{\sigma}^\mu \partial_\mu) \xi_n\ +\ 
\bar{\eta}_n ( i\bar{\sigma}^\mu \partial_\mu) \eta_n\
-\ \frac{n}{R}\, \Big(\, \xi_n \eta_n\, +\, 
\bar{\xi}_n\bar{\eta}_n\,\Big)\nonumber\\            
&&+\,\sqrt{2}\, \Big(\, i\bar{h}^{(n)}_1\, L\tilde{\Phi}_1 \xi_n\ +\
i\bar{h}^{(n)}_2\, L\tilde{\Phi}_2 \eta_n\ +\ 
{\rm H.c.}\,\Big)\, \bigg]\, ,
\end{eqnarray}
where  $\bar{h}^{(n)}_1$  and   $\bar{h}^{(n)}_2$  are given by  Eqs.\ 
(\ref{h1n}) and (\ref{h2n}), respectively.  Observe that the effective
Lagrangian in Eq.\ (\ref{LKK2})  still preserves the original discrete
symmetry  in  Eq.\ (\ref{symm}), where  the KK  components of $N(x,y)$
transform as:  $\xi_n \to i\xi_n$  and $\eta_n \to  -i\eta_n$.  At the
tree  level, the model predicts an  infinite number of KK Dirac states
that have masses whose masses are equally spaced by an interval $1/R$.
Once radiative  corrections are included, however,  as  shown in Fig.\ 
\ref{f1}, each KK Dirac state splits  into a pair of nearly degenerate
Majorana neutrinos.  In fact, radiative  effects induce new  UV-finite
kinetic terms involving  the KK states.  The  new KK kinetic terms are
given by
\begin{equation}
  \label{Lrad}
{\cal L}_{\rm rad}\ =\ \sum_{n,m=1}^\infty\, \kappa_{nm}\, 
\bar{\eta}_n  ( i\bar{\sigma}^\mu \partial_\mu) \xi_m\ +\ {\rm H.c.},
\end{equation}
where simple dimensional analysis of  the Feynman graphs displayed  in
Fig.\   \ref{f1} suggests\footnote{We    should      remark that   our
  renormalization procedure consists of two steps.  In the first step,
  all UV infinities   are absorbed by  off-diagonal wave-function  and
  mixing  renormalizations of  the  KK states in  the  on-shell scheme
  \cite{KP}.  To leading order, such   a rescaling does not  generally
  affect  the original form of   the tree-level effective action.  The
  second step, which is of  interest to us  here, consists of a finite
  renormalization of the kinetic terms.}
\begin{equation}
  \label{kappanm}
\kappa_{nm}\ \sim\ \frac{\bar{h}^{(n)*}_2 \bar{h}^{(m)}_1}{8\pi^2}\ 
\bigg[\, \frac{m^2_{12}}{m^2_{11} + m^2_{22}}\ +\
\frac{m^2_{13}m^{*2}_{23} }{m^2_{33}\, (m^2_{11} + m^2_{22})}\, \bigg]\,.  
\end{equation}
Since   $\kappa_{nm}  \ll 1/(R   M_F)$,  we  find   that,  to  a  good
approximation, only    the diagonal kinetic   transitions  $\xi_n  \to
\eta_n$ contribute predominantly to the splitting  of a KK Dirac state
into a pair of Majorana states.  After  canonically normalizing the KK
kinetic terms, the KK mass spectrum is determined by
\begin{equation}
  \label{LmKK2}
-{\cal L}^{\rm KK}_{\rm mass}\ =\   \sum_{n=1}^{\infty}\,\frac{n}{2R}\ 
\Big(\, \chi^{(n)}_1 \,,\ \chi^{(n)}_2\, \Big)
\left(\! \begin{array}{cc}
\frac{\displaystyle 1}{\displaystyle 1 + |\epsilon_n|} & 0 \\ 0 & 
\frac{\displaystyle 1}{\displaystyle 1 - |\epsilon_n|}  \end{array} \right)
\left( \begin{array}{c} \chi^{(n)}_1 \\ \chi^{(n)}_2 \end{array}
                                           \right)\ +\ {\rm H.c.},
\end{equation}
where $\epsilon_n \approx \kappa_{nn}$ and 
\begin{equation}
  \label{chi12}
\left( \begin{array}{c} \xi_n \\ \eta_n  \end{array}\right)
\ =\  \frac{1}{\sqrt{2}}\, 
\left( 
\begin{array}{cc} e^{-i\phi^n_\epsilon/2} & -e^{-i\phi^n_\epsilon/2} \\ 
e^{i\phi^n_\epsilon/2} & e^{i\phi^n_\epsilon/2} \end{array}\right)\
\left( \begin{array}{cc} \frac{\displaystyle 1}{\displaystyle
\sqrt{1+ |\epsilon_n| }} & 0 \\
       0 & \frac{\displaystyle 1}{\displaystyle \sqrt{1- |\epsilon_n| }}
 \end{array} \right)\   
\left( \begin{array}{c} \chi^{(n)}_1 \\ \chi^{(n)}_2 \end{array}\right)\, ,
\end{equation}
with    $\phi^n_\epsilon  =  {\rm   arg}\,(\epsilon_n)$.   From   Eq.\ 
(\ref{chi12}), we readily see that the radiatively-induced KK Majorana
states, $\chi^{(n)}_1$ and   $\chi^{(n)}_2$,  mix strongly  with   one
another, thus  forming  a  two-level  CP-violating  system,   namely a
CP-violating resonator.  The striking feature  of the present scenario
is  that both the lifting of  the dangerous mass  degeneracy of the KK
Majorana states and  CP violation occurs through  loop  effects.  This
model of   leptogenesis has no  analogue in   4 dimensions,  since the
inclusion  of  an explicit heavy Majorana   mass  is theoretically not
necessary.

\subsection{The hybrid leptogenesis model}

We shall now consider a model based on the  two scenarios discussed in
Sections 3.1 and  3.2, in which  we include the fermionic bilinears $m
\bar{N}  N$ and $M   N^T  C^{(5)-1} N$, as   well  as the three  Higgs
doublets $\Phi_1$, $\Phi_2$ and  $\Phi_3$. As opposed to  the previous
two cases, we shall not impose the $Z_2$ discrete symmetry on the bulk
Dirac  neutrino $N(x,y)$.  As we  will  see, the absence  of the $Z_2$
symmetry yields a distinct prediction for the  mass spectrum of the KK
states.  In particular, we find that the heavy mass scales $m$ and $M$
neither decouple completely from the KK mass spectrum nor get replaced
by other small quantities of order $1/R$.

The effective Lagrangian of the hybrid model reads
\begin{eqnarray}
  \label{Leff3}
{\cal L}_{\rm eff}  & =&  \int\limits_0^{2\pi R}\!\! dy\
 \bigg\{\, \bar{N} \Big( i\gamma^\mu \partial_\mu\, +\, 
 i\gamma_4 \partial_y \Big) N\ -\ m \bar{N} N\ -\ 
\frac{1}{2}\,\Big( M N^T C^{(5)-1} N\ +\
 {\rm H.c.} \Big) \nonumber\\
&& +\, \delta (y-a)\, 
\Big[\, \bar{h}_1\, L\tilde{\Phi}_1 \xi\, +\,
\bar{h}_2\, L \tilde{\Phi}_2 \eta\ +\ {\rm H.c.}\,\Big]\ +\ 
\delta (y-a)\, \Big[\, {\cal L}'_{\rm SM} (\Phi_1 )\,\nonumber\\
&& +\, {\cal L}_V (\Phi_1,\Phi_2,\Phi_3)\,\Big]\, \bigg\}\, .
\end{eqnarray}
The above  Lagrangian  possesses   a  global symmetry given   by  Eq.\ 
(\ref{symm}) which is  only broken softly by  the Higgs mass  terms of
Eq.\ (\ref{m2ij}) and by $M N^T C^{(5)-1} N$.   This is a crucial fact
that ensures the renormalizability of the model.

Since periodicity is the only constraint that applies to $N(x,y)$, the
5-dimensional  two-component spinors   $\xi$ and $\eta$   may then  be
expressed in terms of a Fourier series expansion as follows:
\begin{eqnarray}
  \label{xi3}
\xi (x,y) &=& \frac{1}{\sqrt{2\pi R}}\ \sum_{n=-\infty}^\infty\, \xi_n (x)\ 
                               \exp\bigg(\,\frac{iny}{R}\,\bigg)\,,\\
  \label{eta3}
\eta (x,y) &=& \frac{1}{\sqrt{2\pi R}}\ \sum_{n=-\infty}^\infty\, \eta_n (x)\ 
                               \exp\bigg(\,\frac{iny}{R}\,\bigg)\,.
\end{eqnarray}
Substituting  Eqs.\ (\ref{xi3}) and   (\ref{eta3}) into  the effective
Lagrangian (\ref{Leff3}), we find after $y$ integration
\begin{eqnarray}
  \label{LKK3}
{\cal L}_{\rm eff} & = & {\cal L}'_{\rm SM} (\Phi_1)\ +\ 
{\cal L}_V(\Phi_1,\Phi_2,\Phi_3)\ +\ {\cal L}_{\rm rad}\ 
+\ \sum_{n=-\infty}^\infty\, \bigg\{\, \bar{\xi}_n 
( i\bar{\sigma}^\mu \partial_\mu) \xi_n\ +\ 
\bar{\eta}_n ( i\bar{\sigma}^\mu \partial_\mu) \eta_n\nonumber\\
&&-\, \Big[\, \Big( m + \frac{in}{R}\, \Big)\, \xi_n \eta_{-n}\ +\
{\rm H.c.}\, \Big]\ -\ \frac{1}{2}\, M\, 
\Big( \xi_{-n}\xi_n\, +\, \bar{\eta}_{-n}\bar{\eta}_n\
+\ {\rm H.c.}\Big)\nonumber\\
&& +\, \Big(\, \bar{h}^{(n)}_1\, L\tilde{\Phi}_1 \xi_n\ +\
\bar{h}^{(n)}_2\, L\tilde{\Phi}_2 \eta_n\ +\ 
{\rm H.c.}\,\Big)\, \bigg\}\, ,
\end{eqnarray}
where
\begin{equation}
  \label{h12n}
\bar{h}^{(n)}_1\ =\  \frac{M_F}{M_{\rm P}}\ h_1\,
\exp\bigg(\,\frac{ina}{R}\,\bigg)\,\,,\qquad 
\bar{h}^{(n)}_2\ =\ \frac{M_F}{M_{\rm P}}\  h_2\,
\exp\bigg(\,\frac{ina}{R}\,\bigg)\, .  
\end{equation}
In Eq.\  (\ref{LKK3}),  ${\cal L}_{\rm rad}$  indicates  the UV-finite
radiative contributions to the KK kinetic terms, i.e.
\begin{equation}
  \label{Lrad1}
{\cal L}_{\rm rad}\ =\ \sum_{n,m=-\infty}^\infty\, \kappa_{n,m}\, 
\bar{\eta}_n  ( i\bar{\sigma}^\mu \partial_\mu) \xi_m\ +\ {\rm H.c.},
\end{equation}
where $\kappa_{n,m}$  is given   by a formula   very similar  to  Eq.\ 
(\ref{kappanm}). To avoid  excessive complication  in the calculation,
we consider   only those radiative     terms $\kappa_{n,m}$ that   are
expected  to have  a dominant effect  on  the KK  mass spectrum.  More
explicitly, we have
\begin{eqnarray}
  \label{LKKrad}
{\cal L}_{\rm rad} &\approx& \kappa_{0,0}\, \bar{\eta}_0 
( i\bar{\sigma}^\mu \partial_\mu) \xi_0\ +\ \sum_{n=1}^\infty \Big[\,
\kappa_{n,n}\, \bar{\eta}_n ( i\bar{\sigma}^\mu \partial_\mu) \xi_n\ +\
\kappa_{-n,-n}\, \bar{\eta}_{-n} ( i\bar{\sigma}^\mu \partial_\mu)
\xi_{-n}\nonumber\\
&&+\, \kappa_{n,-n}\, \bar{\eta}_{n} ( i\bar{\sigma}^\mu \partial_\mu)
\xi_{-n}\ +\ \kappa_{-n,n}\, \bar{\eta}_{-n} ( i\bar{\sigma}^\mu \partial_\mu)
\xi_n\, \Big] \quad +\quad {\rm H.c.}
\end{eqnarray}
Notice  that all $|\kappa_{n,m}|$ have  the same absolute value and do
not depend on the indices  $n$ and $m$.  

To evaluate the masses of the KK neutrino states,  it is convenient to
write the kinetic part of the KK sector as a sum of two terms:
\begin{equation}
  \label{LKKkin}
{\cal L}^{\rm KK}_{\rm kin}\ =\ {\cal L}^{\rm KK}_{n = 0}\ +\
{\cal L}^{\rm KK}_{n \ge 1}\,, 
\end{equation}
where 
\begin{eqnarray}
  \label{LKK0}
{\cal L}^{\rm KK}_{n = 0} & =& \Big( \bar{\xi}_0,\ \bar{\eta}_0 \Big)
( i\bar{\sigma}^\mu \partial_\mu)
\left( \begin{array}{cc}
1 & \kappa_{0,0} \\
\kappa_{0,0} & 1 \end{array} \right)\! 
\left( \begin{array}{c} \xi_0 \\ \eta_0 \end{array} \right)\ -\
  \frac{1}{2}
\Big( \xi_0,\ \eta_0 \Big)
\left( \begin{array}{cc}
M & m \\
m & M \end{array} \right)\! 
\left( \begin{array}{c} \xi_0 \\ \eta_0 \end{array} \right)\nonumber\\
&&+\qquad {\rm H.c.}\,,\\
  \label{LKKn}
{\cal L}^{\rm KK}_{n \ge 1} &=& \sum_{n = 1}^\infty\,\Bigg[\ 
\Big( \bar{\xi}_n,\ \bar{\eta}_n,\ \bar{\xi}_{-n},\ \bar{\eta}_{-n} \Big)
( i\bar{\sigma}^\mu \partial_\mu)
\left( \begin{array}{cccc}
1 & \kappa^*_{n,n} & 0 & \kappa^*_{n,-n} \\
\kappa_{n,n} & 1 & \kappa_{n,-n} & 0  \\
0 & \kappa^*_{n,-n} & 1 & \kappa^*_{-n,-n} \\
\kappa_{n,-n} & 0 & \kappa_{-n,-n} & 1  \end{array} \right)\! 
\left( \begin{array}{c} \xi_n\\ \eta_n \\ \xi_{-n}\\ \eta_{-n}
 \end{array} \right)\nonumber\\
&& -\,  \frac{1}{2}\,
\Big( \xi_n,\ \eta_n,\ \xi_{-n},\ \eta_{-n} \Big)
\left( \begin{array}{cccc}
0 & 0 & M & \tilde{m}_n \\
0 & 0 & \tilde{m}^*_n & M \\
M & \tilde{m}^*_n & 0 & 0 \\
\tilde{m}_n & M & 0 & 0 \end{array} \right)\! 
\left( \begin{array}{c} \xi_n\\ \eta_n \\ \xi_{-n}\\ \eta_{-n}
 \end{array} \right)\quad +\quad {\rm H.c.}\,\Bigg]\,,
\end{eqnarray}
with $\tilde{m}_n = m + (in/R)$ (i.e.\ $\tilde{m}_0  = m$). {}From the
Lagrangian  ${\cal L}^{\rm  KK}_{n =  0}$  in  Eq.\  (\ref{LKK0}), one
obtains  two   Majorana neutrinos, $\chi^{(0)}_1$  and $\chi^{(0)}_2$,
with masses
\begin{equation}
  \label{mchi0}
m^{(0)}_{\chi_1}\ =\ \frac{|M-m|}{|1-\kappa_{0,0}|}\ \approx\ |M-m|,\qquad
m^{(0)}_{\chi_2}\ =\ \frac{M+m}{|1+\kappa_{0,0}|}\ \approx\ M+m\,. 
\end{equation}
We now turn to the evaluation of the KK  neutrino masses, for the more
involved case with $n\ge 1$. To this end, we first go  to a weak basis,
in which the mass matrix is real, by rephasing the KK fields:
\begin{equation}
  \label{phin}
\xi_n\ \to\ e^{-i\phi_n/2}\, \xi_n\,,\quad
\eta_n\ \to\ e^{i\phi_n/2}\, \eta_n\,,\quad
\xi_{-n}\ \to\ e^{i\phi_n/2}\, \xi_{-n}\,,\quad
\eta_{-n}\ \to\ e^{-i\phi_n/2}\, \eta_{-n}\,,  
\end{equation}
with $\phi_n = {\rm arg}\, \tilde{m}_n$.  Even though one could always
work out the most  general case, it  is, however, very illuminating to
make a further assumption that leads to much simpler analytic results.
We  assume that all radiative  kinetic terms are predominantly real in
the new weak basis in Eq.\ (\ref{phin}), i.e.\ ${\rm Im}\,\kappa_{n,m}
\ll {\rm Re}\,\kappa_{n,m} \approx \epsilon$.  Then, considering $m\ge
M$   for definiteness, we  can    diagonalize the Lagrangian in   Eq.\ 
(\ref{LKKn}) through the canonical transformation
\begin{equation}
  \label{cantr}
\left( \begin{array}{c} \xi_n\\ \eta_n \\ \xi_{-n}\\ \eta_{-n}
 \end{array} \right)\ =\ \frac{1}{2}\, 
\left( \begin{array}{cccc}
i & 1 & -i & 1\\
-i & -1 & -i & 1\\
-i & 1 & i & 1 \\
i & -1 & i & 1 \end{array} \right)
\left( \begin{array}{cccc}
i & 0 & 0 & 0\\
0 & \frac{\displaystyle i}{\displaystyle \sqrt{1-2\epsilon}} & 0 & 0\\
0 & 0 & 1 & 0 \\
0 & 0 & 0 & \frac{\displaystyle 1}{\displaystyle \sqrt{1+2\epsilon}} 
\end{array} \right)
\left( \begin{array}{c} \chi^{(n)}_1 \\ \chi^{(n)}_2 \\ 
\chi^{(-n)}_1 \\ \chi^{(-n)}_2 \end{array} \right)\, ,
\end{equation}
which leads to the KK Majorana masses 
\begin{eqnarray}
  \label{mKKn}
m^{(n)}_{\chi_1} \!&=&\! \sqrt{m^2 + \frac{n^2}{R^2}}\, -\, M\ ,\qquad
m^{(n)}_{\chi_2}\ =\ \frac{1}{1-2\epsilon}\ \Bigg(\, 
\sqrt{m^2 + \frac{n^2}{R^2}}\, -\, M\, \Bigg)\, ,\nonumber\\
m^{(-n)}_{\chi_1} \!&=&\! M\, +\, \sqrt{m^2 + \frac{n^2}{R^2}}\ ,\qquad
\!\!\!m^{(-n)}_{\chi_2}\ =\ \frac{1}{1+2\epsilon}\ \Bigg(\, 
M\, +\, \sqrt{m^2 + \frac{n^2}{R^2}}\ \ \Bigg)\, .
\end{eqnarray}
Evidently, Eq.\   (\ref{mKKn})   shows that the  immediate   effect of
radiative corrections is to lift the dangerous twofold mass degeneracy
among the KK  states $\chi^{(\pm n)}_1$  and $\chi^{(\pm  n)}_2$, thus
rendering the theory CP-violating.  If one considers that $m > M$, the
mass of  the lowest-lying KK  state is $m^{(0)}_{\chi_1} \approx m-M$,
which can  naturally be much larger  than the  compactification scale
$1/R$.  This  is  a distinctive  feature of the  present model, since,
unlike the previous two scenarios, the  heavy mass scale $m-M$ neither
decouples  from  the complete  KK mass spectrum   nor gets replaced by
quantities of order $1/R$.

\setcounter{equation}{0}
\section{Resonant CP violation}

In  addition   to   lepton-number    violation,   CP  non-conservation
constitutes another important ingredient for leptogenesis.   These two
necessary conditions satisfy,   by construction, the   three models of
leptogenesis,   discussed in the   previous  section.   However, these
conditions may not be  sufficient to guarantee an appreciable leptonic
asymmetry that results from decays  of KK Majorana states according to
the standard  scenario of leptogenesis  \cite{FY}.   In particular, in
theories with low scale of quantum gravity, we have to ensure that the
total net effect of  the individual CP-violating contributions  coming
from the tower of nearly degenerate KK Majorana states does not vanish
because of some  kind of a GIM  \cite{GIM} cancellation  mechanism. In
fact, by making use of such a GIM-type mechanism, we can show that all
the CP-violating  vertex  ($\varepsilon'$-type) terms  almost   cancel
pairwise.  On  the other  hand, we find  that the  interference of the
CP-violating self-energy ($\varepsilon$-type)       contributions   is
constructive or destructive,  depending on the mass  spacing of the KK
Majorana states.
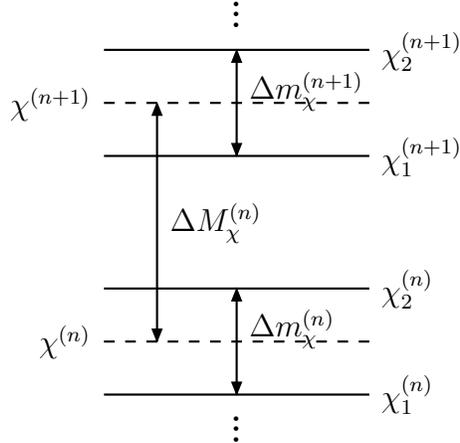
\begin{figure}
\begin{center}
\begin{picture}(100,150)(0,0)
\SetWidth{0.8}

\Line(0,10)(100,10)\Text(105,10)[l]{$\chi^{(n)}_1$}
\DashLine(0,30)(100,30){4}\Text(-5,30)[r]{$\chi^{(n)}$}
\Line(0,50)(100,50)\Text(105,50)[l]{$\chi^{(n)}_2$}

\LongArrow(20,30)(20,119)\LongArrow(20,35)(20,31)
\Text(25,75)[l]{$\Delta M^{(n)}_\chi$}

\Line(0,100)(100,100)\Text(105,100)[l]{$\chi^{(n+1)}_1$}
\DashLine(0,120)(100,120){4}\Text(-5,120)[r]{$\chi^{(n+1)}$}
\Line(0,140)(100,140)\Text(105,140)[l]{$\chi^{(n+1)}_2$}

\LongArrow(50,10)(50,49)\LongArrow(50,15)(50,11)
\Text(55,35)[l]{$\Delta m^{(n)}_\chi$}

\LongArrow(50,100)(50,139)\LongArrow(50,105)(50,101)
\Text(55,125)[l]{$\Delta m^{(n+1)}_\chi$}

\Text(50,157)[]{\bf $\vdots$}\Text(50,0)[]{\bf $\vdots$}

\end{picture}
\end{center}
\caption{Schematic representation of the mass spectrum of the KK 
states.}\label{f2}
\end{figure}

The mass spectrum   of the higher-dimensional models   of leptogenesis
under discussion consists of  an  infinite series  of pairs of  nearly
degenerate Majorana  neutrinos, which we  denote by $\chi^{(n)}_1$ and
$\chi^{(n)}_2$.  The generic  pattern of the KK  mass  spectrum may be
represented by Fig.\ \ref{f2}. As we have discussed  in Section 3, the
mass   difference between  $\chi^{(n)}_1$   and $\chi^{(n)}_2$  may be
induced either at the tree level:
\begin{equation}
  \label{DmKK0}
\Delta m^{(n)}_\chi\ \equiv\ m^{(n)}_{\chi_2}\, -\,  m^{(n)}_{\chi_1}\
=\ 2\mu\, ,
\end{equation}
or through radiative kinetic terms:
\begin{equation}
  \label{DmKK1}
\Delta m^{(n)}_\chi\ \sim\ \kappa_{nm}\, m^{(n)}_\chi\ \sim\
\frac{h^*_2h_1}{8\pi^2}\, \frac{M^2_F}{M^2_{\rm P}}\, m^{(n)}_\chi\, ,
\end{equation}
with
\begin{equation}
  \label{mnchi}
m^{(n)}_\chi\ \equiv\ \frac{1}{2}\ \Big(\, 
m^{(n)}_{\chi_1}\, +\,  m^{(n)}_{\chi_2}\, \Big)\, .  
\end{equation}
Furthermore,   the mass difference  between two  adjacent  KK pairs is
determined by
\begin{equation}
  \label{mpair}
\Delta M^{(n)}_\chi\ \equiv\ 
m^{(n+1)}_\chi\, -\, m^{(n)}_\chi\ \approx\ \frac{1}{R}\ =\
\bigg(\, \frac{M_F}{M_{\rm P}}\, \bigg)^{2/\delta}\, M_F\, .
\end{equation}
In deriving the approximate   equality in Eq.\ (\ref{mpair}),  we have
implicitly assumed that $m \stackrel{<}{{}_\sim}  n/R$, for the hybrid
scenario  outlined in   Section 3.3.  Clearly,  if  the  origin  of  a
non-zero $\Delta  m^{(n)}_\chi$ is due to  loop effects, one naturally
has $\Delta   M^{(n)}_\chi  \gg \Delta m^{(n)}_\chi$  for   any number
$\delta$  of  extra dimensions.    However,  if $\Delta  m^{(n)}_\chi$
occurs in the   Born approximation,  then  $\Delta  m^{(n)}_\chi$  and
$\Delta M^{(n)}_\chi$ could  be of equal order.  As we will see below,
the  last two quantities determine the  magnitude of CP violation that
originates from the interference of the tower of KK states.

Let us  first consider $\varepsilon'$-type  CP violation in the decays
of KK states.  For our illustrations, it is  sufficient to assume that
the KK states decay predominantly to the SM Higgs doublet $\Phi$ or to
the Higgs doublet  with the smallest (thermal)  mass in the model with
the  extended Higgs  sector.    Since $\Delta  m^{(n)}_\chi   < \Delta
M^{(n)}_\chi \ll m^{(n)}_\chi$, the CP-violating parameter of interest
to us is
\begin{equation}
  \label{epsprime}
\varepsilon'^{(n)}_\chi\ =\ \frac{ |{\cal T}^{(n)\varepsilon'}_{\chi_1}|^2
+ |{\cal T}^{(n)\varepsilon'}_{\chi_2}|^2 - |\overline{\cal
  T}^{(n)\varepsilon'}_{\!\!\chi_1}|^2 - |\overline{\cal
  T}^{(n)\varepsilon'}_{\!\!\chi_2}|^2 }{ 
|{\cal T}^{(n)\varepsilon'}_{\chi_1}|^2
+ |{\cal T}^{(n)\varepsilon'}_{\chi_2}|^2 + |\overline{\cal
  T}^{(n)\varepsilon'}_{\!\!\chi_1}|^2 + |\overline{\cal
  T}^{(n)\varepsilon'}_{\!\!\chi_2}|^2   }\ ,
\end{equation}
where we used the short-hand notations  for the transition amplitudes:
${\cal  T}^{(n)\varepsilon'}_{\chi_1} =    {\cal  T}^{\varepsilon'}  (
\chi^{(n)}_1       \to    L       \Phi^\dagger   )$,   $\overline{\cal
  T}^{(n)\varepsilon'}_{\!\!\chi_1}  = {\cal    T}^{\varepsilon'}    (
\chi^{(n)}_1 \to L^C \Phi )$, and likewise for $\chi^{(n)}_2$.  In all
these amplitudes, only vertex diagrams  are included.  For simplicity,
we assume  that all Yukawa couplings  $h^{(k)}_1$  and $h^{(k)}_2$ are
independent of $k$, although the most general  case does not depend on
that particular  assumption.  The parameter  $\varepsilon'^{(n)}_\chi$
is then found to be
\begin{equation}
  \label{epsKKprime}
\varepsilon'^{(n)}_\chi\ =\ \frac{1}{8\pi\, ( |h^{(n)}_1|^2 +
|h^{(n)}_2|^2 )}\  \sum_k\,  {\rm Im}\, 
(h^{(n)*}_1h^{(k)}_2 )^2\ \Bigg[\, 
f\Bigg(\,\frac{m^{(k)\,2}_{\chi_2}}{m^{(n)\,2}_{\chi_1} } \Bigg)\ -\
f\Bigg(\,\frac{m^{(k)\,2}_{\chi_1}}{m^{(n)\,2}_{\chi_2} } \Bigg)\ \Bigg]\,,
\end{equation}
with 
\begin{equation}
  \label{fx}
f(x)\ =\ \sqrt{x}\, \bigg[\, 1\, -\, \Big( 1 + x\Big)\, 
\ln\bigg( 1 +\frac{1}{x}\bigg)\, \bigg]\, .  
\end{equation}
Note that the range of summation over the KK states explicitly depends
on the model.  Equation (\ref{epsKKprime}) may further be approximated
as
\begin{equation}
  \label{epappr}
\varepsilon'^{(n)}_\chi\ \approx\ \frac{1}{4\pi\, ( |h^{(n)}_1|^2 +
|h^{(n)}_2|^2 )}\  \sum_k\,  {\rm Im}\, 
(h^{(n)*}_1h^{(k)}_2 )^2\, 
\Bigg(\, \frac{\Delta m^{(n)}_\chi}{m^{(n)}_\chi}\, +\,
\frac{\Delta m^{(k)}_\chi}{m^{(k)}_\chi}\, \Bigg)\, 
\frac{m^{(k)\,2}_\chi}{m^{(n)\,2}_\chi}\,
f'\Bigg(\,\frac{m^{(k)\,2}_\chi}{m^{(n)\,2}_\chi } \Bigg)\, ,
\end{equation}
where $f'(x)$ is the derivative of the function $f(x)$, i.e.
\begin{equation}
  \label{fx'}
f'(x)\ =\ \frac{3}{2\sqrt{x}}\ \bigg[\, 1\, -\, 
\bigg(\, \frac{2}{3} + x\bigg)\, 
\ln\bigg( 1 +\frac{1}{x}\bigg)\, \bigg]\, .  
\end{equation}
It is  obvious that each individual  KK term in Eq.\ (\ref{epappr}) is
suppressed  by a factor  $\Delta m^{(k)}_\chi/m^{(k)}_\chi$. This is a
generic consequence of a  GIM-type cancellation mechanism, as a result
of off-shell interference between pseudo-Dirac neutrinos.

To explicitly demonstrate that $\varepsilon'$-type contributions to CP
violation are indeed small, it is instructive to offer an estimate for
the sum  over the KK states  in Eq.\  (\ref{epappr}), after making few
plausible assumptions.  For simplicity, we consider  a theory with one
additional spatial dimension ($\delta  =1$),  and further assume  that
the mass differences,  $\Delta m^{(k)}_\chi$ (or $\Delta m^{(k)}_\chi/
m^{(k)}_\chi$) and $\Delta M^{(k)}_\chi$,  are independent of $k$.  In
addition, we convert  the sum over the  KK  states `$k$' to  an energy
integral, which  has a UV  cutoff at the  fundamental scale of quantum
gravity $M_F$.  With these considerations, we obtain
\begin{equation}
  \label{epappr1}
\varepsilon'^{(n)}_\chi\ \approx\ -\, \frac{{\rm Im}\, 
(h^{(n)*}_1h^{(n)}_2 )^2}{16\pi\, ( |h^{(n)}_1|^2 +
|h^{(n)}_2|^2 )}\  
\frac{\Delta m^{(n)}_\chi}{\Delta M^{(n)}_\chi}\
\approx\ -\, \frac{{\rm Im}\, 
(h^*_1h_2 )^2}{16\pi\, ( |h_1|^2 +
|h_2|^2 )}\ \frac{M^2_F}{M^2_{\rm P}}\ 
\frac{\Delta m^{(n)}_\chi}{\Delta M^{(n)}_\chi}\  ,
\end{equation}
where  we have used the fact  that  $f'(x) \approx -1/(2x^{3/2})$, for
$x\gg 1$.  The above exercise  shows that $\varepsilon'^{(n)}_\chi$ is
extremely   suppressed by    the    extra-dimensional   volume  factor
$M^2_F/M^2_{\rm P}$  and, in certain scenarios,  by  the ratio $\Delta
m^{(n)}_\chi  / \Delta  M^{(n) }_\chi \ll  1$.   Consequently, we  can
safely  neglect $\varepsilon'$-type CP  violation in the decays of the
KK Majorana states.

In the  following, we shall   focus our attention on  the  self-energy
($\varepsilon$-type) contribution to CP violation.  In analogy to Eq.\ 
(\ref{epsprime}), the    relevant  measure of    $\varepsilon$-type CP
violation may be defined by
\begin{equation}
  \label{epsilon}
\varepsilon^{(n)}_\chi\ =\ \frac{ |{\cal T}^{(n)\varepsilon}_{\chi_1}|^2
+ |{\cal T}^{(n)\varepsilon}_{\chi_2}|^2 - |\overline{\cal
  T}^{(n)\varepsilon}_{\!\!\chi_1}|^2 - |\overline{\cal
  T}^{(n)\varepsilon}_{\!\!\chi_2}|^2 }{ 
|{\cal T}^{(n)\varepsilon}_{\chi_1}|^2
+ |{\cal T}^{(n)\varepsilon}_{\chi_2}|^2 + |\overline{\cal
  T}^{(n)\varepsilon}_{\!\!\chi_1}|^2 + |\overline{\cal
  T}^{(n)\varepsilon}_{\!\!\chi_2}|^2   }\ .
\end{equation}
Correspondingly, ${\cal T}^{(n)\varepsilon}_{\chi_{1,2}}$ ($\overline{
  \cal T}^{(n)  \varepsilon}_{\!\!\chi_{1,2}}$)   indicate the  decays
$\chi^{(n)}_{1,2}\to  L\Phi^\dagger$ ($\chi^{(n)}_{1,2}\to L^C \Phi$),
where only the self-energy  graph has been taken  into account.  Since
each $\varepsilon$-type term is proportional to the mass difference of
the KK states involved, one  has to avoid that self-energy transitions
$\chi^{(n+1)}_1  \to  \chi^{(n+1)}_2$  cancel  against the transitions
$\chi^{(n+1)}_1   \to \chi^{(n)}_2$.   From   Fig.\  \ref{f2}, we  may
schematically deduce the condition  for destructive interference among
adjacent KK states, which translates into the relation
\begin{equation}
  \label{destr}
\Delta m^{(n)}_\chi\ \sim\ \frac{1}{2}\, \Delta M^{(n)}_\chi\, .
\end{equation}
Instead, if
\begin{equation}
  \label{constr}
\Delta M^{(n)}_\chi\ \gg\ \Delta m^{(n)}_\chi\, ,
\end{equation}
the interference  of two   neighbouring KK  states  is  constructive.  
Employing   the resummation approach   to    the mixing of    unstable
particles, which was developed in \cite{APRD,PP}, we find that
\begin{equation}
  \label{eps}
\varepsilon^{(n)}_\chi\ =\ \frac{2\,{\rm Im}\, (h^{(n)}_1 
h^{(n)*}_2 )^2 }{ ( |h^{(n)}_1|^2 + |h^{(n)}_2|^2)^2}\
\frac{\Delta m^{(n)}_\chi\,
( \Gamma^{(n)}_{\chi_1}\, +\, \Gamma^{(n)}_{\chi_2} )}{
(\Delta m^{(n)}_\chi )^2 + \frac{1}{4}\Gamma^{(n)2}_{\chi_2} }\ 
\Bigg[\, 1\, +\, \frac{(\Delta m^{(n)}_\chi )^2 + \frac{1}{4}
 \Gamma^{(n)2}_{\chi_2}}{ (\Delta m^{(n)}_\chi )^2 + \frac{1}{4}
\Gamma^{(n)2}_{\chi_1} }\, \Bigg]\,,
\end{equation}
where 
\begin{equation}
  \label{GammaKK}
\Gamma^{(n)}_{\chi_1}\ =\ \frac{1}{8\pi}\ |h^{(n)}_1|^2\,
m^{(n)}_{\chi_1}\quad {\rm and}\quad
\Gamma^{(n)}_{\chi_2}\ =\ \frac{1}{8\pi}\ |h^{(n)}_2|^2\,
m^{(n)}_{\chi_2}
\end{equation}
are the     decay widths     of $\chi^{(n)}_1$   and   $\chi^{(n)}_2$,
respectively.  In Eq.\ (\ref{eps}), we have neglected contributions to
$\varepsilon^{(n)}_\chi$  of    order $\Delta  m^{(n)}_\chi  /  \Delta
M^{(n)}_\chi$ (cf.\ Eq.\   (\ref{constr})) and $\Delta m^{(n)}_\chi  /
m^{(n)}_\chi$.

In agreement with Ref.\  \cite{APRD}, we observe that the CP-violating
parameter $\varepsilon^{(n)}_\chi$ given by Eq.\ (\ref{eps}) can be of
order 1, if the two conditions:
\begin{equation}
  \label{CPres}
{\rm (i)}\ \delta^{(n)}_{\rm CP}\ \equiv\ 
\frac{|{\rm Im}\, (h^{(n)}_1 
h^{(n)*}_2 )^2| }{ |h^{(n)}_1|^2 |h^{(n)}_2|^2}\ \sim\
1\quad {\rm and}\quad
{\rm (ii)}\  \Delta m^{(n)}_\chi\ \sim\
\frac{\Gamma^{(n)}_{\chi_1}}{2}\ \,{\rm or}\,\ 
\frac{\Gamma^{(n)}_{\chi_2}}{2} 
\end{equation}
are satisfied. The first condition is rather model-dependent. A priori
there is no  reason to believe that the  phases of the original Yukawa
couplings $h_1$ and $h_2$  should somehow be aligned  and, as a result
of  this, the parameter  $\delta^{(n)}_{\rm CP}$  must  be suppressed. 
Thus, we consider $\delta^{(n)}_{\rm CP} \sim 1$.

In  a general model, it  is more  difficult, however, to theoretically
justify the  second condition,  as  the mass   splitting of the  mixed
particles  involved  must be of the  order  of their widths.   For the
scenarios discussed   in Sections  3.2 and  3.3,  the  mass  splitting
$\Delta m^{(n)}_\chi$ is radiatively  induced  by integrating out  the
Higgs interactions  and  then  canonically normalizing  the  resulting
kinetic terms.  Thus, the  width of the KK  Majorana states  and their
respective mass difference formally occur at the same electroweak loop
order.  Therefore, the  second condition is  naturally implemented for
these two models.   For  the leptogenesis model described  in  Section
3.1,  one  has  to assume   that $\mu \sim   \Gamma^{(n)}_{\chi_1}$ or
$\Gamma^{(n)}_{\chi_2}$.  As a  consequence of compactification of the
extra large  dimensions, the mass parameter $\mu$  always turns out to
be smaller than $\Delta M^{(n)}_\chi$, so  some degree of tuning $\mu$
to even smaller values is required in this case.

The models of leptogenesis   we have been  studying share  the generic
feature that  each KK Dirac  state decomposes  into  a pair of  nearly
degenerate  Majorana neutrinos.  Such a  pair of KK Majorana neutrinos
forms a strongly mixed two-level  system that exhibits CP violation of
order unity; such a  system was called  a CP-violating resonator.   In
fact, it was shown in \cite{APRD}  that the resonant enhancement of CP
violation is driven  by  the non-diagonalizable (Jordan-like)  form of
the effective Hamiltonian (or equivalently the resummed propagator) of
the  two-level system,  which satisfies  conditions  very analogous to
those   of  Eq.\ (\ref{CPres}).  Finally,   we  should stress that the
constructive   interference   of all  the individual   KK CP-violating
resonators  is assured on the  basis of the  requirement given by Eq.\ 
(\ref{constr}).  This last requirement  is more naturally  implemented
in the models of  leptogenesis  with extended  Higgs sector (see  also
discussion in Sections 3.2 and 3.3).

\setcounter{equation}{0}
\section{Baryonic asymmetry of the Universe}

Astronomical    observations  give strong    evidence that the present
Universe mainly consists of  matter rather than antimatter,  i.e.\ the
Universe  possesses  an excess in the  $B$   number.  The observed $B$
asymmetry  may be quantified  by  the nonzero baryon-number-to-entropy
ratio of densities \cite{KT}
\begin{equation}
  \label{nDB}
Y_{\Delta B}\ =\ \frac{n_{\Delta B}}{s}\ =\  (0.6-1)\times 10^{-10}\ ,  
\end{equation}
where $n_{\Delta B} =  n_B - n_{\bar{B}}  \approx n_B$ and $s$  is the
entropy density.  As we  mentioned in the introduction,  an attractive
solution that could account for the nonzero value of $Y_{\Delta B}$ by
making     use all of the    necessary  conditions imposed by Sakharov
\cite{ADS} may  be given  by means of   the scenario  of  baryogenesis
through leptogenesis  \cite{FY}.  Based  on  an analysis   of chemical
potentials \cite{HT}, one may derive that
\begin{equation}
  \label{BLrel}
Y_{\Delta B} (T > T_c) \ =\ \frac{8N_F + 4N_H}{22N_F + 13 N_H}\ 
Y_{\Delta (B - L)}\ .  
\end{equation}
Almost independently of  the numbers $N_F$  and $N_H$ of  flavours and
Higgs doublets, one finds  that approximately one third of the initial
$B-L$ and/or $L$ asymmetry will  be reprocessed  into an asymmetry  in
$B$, provided sphalerons are  in  thermal equilibrium.  If the  reheat
temperature  $T_r$ is  smaller than  the  critical temperature  $T_c$,
sphalerons are out of equilibrium, and the above $L$-to-$B$ conversion
will be  exponentially    suppressed by a   factor  $\exp  (-T_c/T_r)$
\cite{KT}.

Let us first consider the  constraints on the  parameter space of  the
leptogenesis models,  coming from Sakharov's  requirement  that $L$ or
$B-L$-violating  processes, such as  decays of KK Majorana modes, must
be  out-of-thermal equilibrium  in  an  expanding  Universe.   As  was
discussed by Abel and  Sarkar \cite{AS}, the  presence of low-lying KK
states drastically influences  the evolution of  the Universe,  as the
number of relativistic degrees  of freedom increases with  temperature
$T$.  To  be more  precise, if  $m^{(0)}_{\chi_1} \equiv m_{\rm  min}$
represents  the mass of  the   lowest KK  state in   a given model  of
leptogenesis, the  number of relativistic KK states  below $T$ is then
roughly  given by  $[(T-m_{\rm min})R]^\delta$, where  $\delta$ is the
number  of  large  compact  dimensions.   Thus, the  number of  active
degrees of freedom at a given temperature $T$ is determined by
\begin{eqnarray}
  \label{gT}
g (T) &\approx& g_*\, +\, S_\delta\,\theta(T-m_{\rm min}) \,
[(T-m_{\rm min})R]^\delta\nonumber\\ 
&\approx& g_*\, +\, S_\delta\,\theta(T-m_{\rm min})\, 
\bigg(\frac{M_{\rm P}}{M_F}\bigg)^2\, 
\bigg(\frac{T-m_{\rm min} }{M_F}\bigg)^\delta\, ,
\end{eqnarray}
where $g_*\approx  100$ is the number  of active degrees of freedom in
usual   4-dimensional   extensions of   the     SM, and  $S_\delta   =
2\pi^{\delta/2} / \Gamma(\frac{\delta}{2})$ is  the surface area  of a
$\delta$-dimensional sphere of  unit radius.  {}From  Eq.\ (\ref{gT}),
we see that the part of $g(T)$ modified by  the presence of KK states,
$g_{\rm   KK} (T)$, can generally  be  much larger  than $g_*$, unless
$m_{\rm  min} \sim T_c$, or  $M_F$ and $\delta$  are sufficiently high
for  some  specific model.   For instance, in  the hybrid leptogenesis
model, one may have $m_{\rm min} \approx m - M > T_c$, and $g_{\rm KK}
(T)$ is of order $g_*$ for $T \stackrel{>}{{}_\sim} T_c$.

Sakharov's   requirement  that  all  $B$-   and,  because  of possible
equilibrated sphaleron interactions, $L$-violating processes should be
out of thermal equilibrium translates  into the approximate inequality
for the total $T$-dependent decay rate of the KK states
\begin{equation}
  \label{EQcond}
\Gamma_\chi (T)\ \equiv 
\sum_{n = {\rm int}\, (m_{\rm min} R )}^{{\rm int}\,(T
  R)}   \Gamma^{(i)}_{\chi} \ \stackrel{<}{{}_\sim}\ 2 H (T)\, ,
\end{equation}
where   $n = (n_1,  n_2, \dots,   n_\delta  )$, $\Gamma^{(n)}_{\chi} =
\frac{1}{2}   (\Gamma^{(n)}_{\chi_1}  + \Gamma^{(n)}_{\chi_2})$ is the
average decay width of the $n$th CP-violating resonator, and
\begin{equation}
  \label{HT}
H(T)\ =\ 1.73\, g^{1/2} (T)\, \frac{T^2}{M_{\rm P}}\ \approx\
1.73\, S^{1/2}_\delta\, \bigg(\frac{T-m_{\rm min}
  }{M_F}\bigg)^{\delta/2}\, \frac{T^2}{M_F}\  .  
\end{equation}
The last approximate equality holds true, provided $g_{\rm KK} (T) \gg
g_*$.  Converting the  multidimensional sum over the  KK modes in Eq.\ 
(\ref{EQcond}) into an integral, we find that
\begin{equation}
  \label{GamT}
\Gamma_\chi (T)\ \approx\ \frac{|h_1|^2 + |h_2|^2}{16\pi^2}\,
\frac{S_\delta}{\delta}\,  \bigg(\frac{T-m_{\rm min}
  }{M_F}\bigg)^{\delta +1}\, M_F\, . 
\end{equation}
An immediate result    of  the out-of-equilibrium  condition in   Eq.\
(\ref{EQcond}) is the constraint
\begin{equation}
  \label{h1h2bound}
\frac{1}{2}\ \Big(|h_1|^2 + |h_2|^2\Big)\ \stackrel{<}{{}_\sim}\ 
32\pi^2\, S^{-1/2}_\delta\, \frac{T^2}{M^2_F}\, \bigg(\frac{T-m_{\rm min}
  }{M_F}\bigg)^{-1-\delta/2}\, ,    
\end{equation}
which  no    longer     depends on    $M_{\rm    P}$.     {}From  Eq.\ 
(\ref{h1h2bound}),    it  is  interesting  to     see that no  serious
arrangement of the parameters is necessary for  all $\delta \ge 1$ and
$m_{\rm min} < T  < M_F$, even  if the original Yukawa couplings $h_1$
and $h_2$  in Eq.\  (\ref{h5h4}) are taken   to be of  order 1.   This
should  be contrasted with  the  extremely tight limits  on the Yukawa
couplings in the conventional 4-dimensional models \cite{APRD}, namely
$h_{1,2} \stackrel{<}{{}_\sim} 10^{-6}$. These  limits are obtained if
one sets $m_{\rm   min} = \delta =  0$,  $T = 0.2$--1  TeV  and $M_F =
M_{\rm  P}$,  and     replaces    $S_\delta$ by  $1/g_*$      in  Eq.\ 
(\ref{h1h2bound}).

It is interesting  to derive the  time evolution of  the Universe as a
function of its temperature in higher-dimensional theories.  We assume
that the Friedmann--Robertson--Walker  metric governs the expansion of
the   Universe   after  inflation  \cite{KT},   and that    all active
relativistic  degrees  of freedom  are  in  chemical  equilibrium  and
therefore have the  same temperature.  Imposing  entropy conservation,
i.e.\ 
\begin{equation}
  \label{sT}
s R^3\ \propto\ g(T)\, T^3 R^3\ =\ {\rm const.},   
\end{equation}
and differentiating with respect to time $t$, we find that 
\begin{equation}
  \label{Htemp}
H\ \equiv\ \frac{1}{R}\, \frac{dR}{dt} 
=\ -\,\frac{\delta + 3}{3}\, \frac{1}{T}\, \frac{dT}{dt}\ ,  
\end{equation}
for $g_{\rm KK} (T) \gg g_*$.  If we differentiate the Hubble variable
in Eq.\ (\ref{HT}) with respect to $t$  and employ Eq.\ (\ref{Htemp}),
we arrive at the differential equation
\begin{equation}
  \label{Hdiff}
\frac{d H}{dt}\ =\ -\, \frac{3}{2}\, \frac{\delta + 4}{\delta + 3}\
H^2\ .  
\end{equation}
Considering as initial condition $H(t\to 0)  \to \infty$, the solution
of Eq.\ (\ref{Hdiff}) reads
\begin{equation}
  \label{tTemp}
t (T)\ =\ \frac{2}{3}\, \frac{\delta + 3}{\delta + 4}\, \frac{1}{H(T)}\ 
\approx\ \Big( 7.6\times 10^{-28}\ {\rm sec}\Big)\ 
\frac{1}{S^{1/2}_\delta}\, \frac{\delta + 3}{\delta + 4}\,
\bigg(\frac{{\rm TeV}}{M_F}\bigg)\,\bigg(\frac{T}{M_F}\bigg)^{- 2 - 
\frac{1}{2}\delta }\, .
\end{equation}
If  $g_{\rm  KK} (T)  \sim  g_*$, which  happens   for temperatures $T
\stackrel{<}{{}_\sim}  m_{\rm    min} +     (M_F/M_{\rm P})^{2/\delta}
g^{1/\delta}_* M_F$,  the time-temperature relation (\ref{tTemp}) goes
over into the canonical 4-dimensional form
\begin{equation}
  \label{tHcan}
t (T)\ =\ \frac{1}{2 H(T)}\ \approx\ (2.3\ {\rm
  sec})\times g^{-1/2}_*\, \bigg(\frac{{\rm MeV}}{T}\bigg)^2\, .  
\end{equation}
{}Comparing   the    $T$-dependences  in     Eqs.\   (\ref{tTemp}) and
(\ref{tHcan}), one readily  sees  that the  presence of  large compact
dimensions drastically changes the evolution of the Universe.

We shall  now attempt to  give an estimate   of the baryonic asymmetry
that results   from  a    sphaleron-converted  leptonic  asymmetry  in
KK-neutrino   decays, with masses  $m^{(n)}_\chi  >  M_\Phi \sim T_c$,
including possible suppression factors due to a low reheat temperature
$T_r$.  As a starting point,  we assume that $n^{(n)}_\chi (T) \approx
n_\gamma   (T)   $,   for   $m^{(n)}_\chi   \stackrel{<}{{}_\sim}    T
\stackrel{<}{{}_\sim} M_F$,   where $n^{(n)}_\chi  (T)$ is the  number
density of  the  $n$th KK pair of   Majorana states and $n_\gamma  (T)
\approx 2.4\, T^3/\pi^2$ is the respective number density for photons.
The dominant   contribution to the  $L$ asymmetry   is expected  to be
encoded in the $n$th CP-violating resonator at $T\approx m^{(n)}_\chi$
for $m^{(n)}_\chi > M_\Phi \sim T_c$ and $m^{(n)}_\chi > m_{\rm min}$,
when the  equilibrium number density  of the $n$th KK  pair is  of the
order of $n^{(n)}_\chi (T)$ \cite{KT}.   Thus, the $n$th  CP-violating
resonator gives rise to a leptonic excess
\begin{equation}
  \label{YLn}
Y^{(n)}_{\Delta L}\ \approx\ \frac{\varepsilon^{(n)}_\chi\, 
n^{(n)}_\chi(m^{(n)}_\chi )}{s (m^{(n)}_\chi )}\ \approx\ 
\frac{\varepsilon^{(n)}_\chi}{ g (m^{(n)}_\chi)}\ .
\end{equation}
In deriving the last  step of Eq.\ (\ref{YLn}),  we have used the fact
that $s (T) \approx g (T)  n_\gamma$.  Since low-scale quantum gravity
theories are plagued by the low-reheat-temperature problem, i.e.\ $T_r
\ll T_c$,  the conversion of  an $L$-to-$B$ asymmetry can only proceed
via sphaleron interactions,  which  are out  of thermal equilibrium.   
Such    an   $L$-to-$B$   conversion   mediated by  out-of-equilibrium
sphalerons may  be taken into account by  multiplying the RHS  of Eq.\ 
(\ref{YLn}) with an  exponentially suppressed factor $\exp (-T_c/T_r)$
\cite{KT}. In this qualitative picture, the total $B$ asymmetry may be
estimated by
\begin{equation}
  \label{YBtot}
Y_{\Delta B}\ \approx\ -\, \frac{1}{3}\, e^{-T_c/T_r}\,
\sum_{n = {\rm int}\, (T_{\rm min} R )}^{{\rm int}\,
(M_F R)}\ Y^{(n)}_{\Delta L}\, .   
\end{equation}
In the    leptogenesis models under   discussion,   all the individual
CP-violating  asymmetries $\varepsilon^{(n)}_\chi$   are of  the  same
order, i.e.\ $\varepsilon^{(n)}_\chi = -\varepsilon_\chi$ for all $n$,
and their net effect is constructive, as long as the condition in Eq.\
(\ref{constr}) is satisfied.  For  generality, let us assume that  the
interference of the CP-violating  resonators is constructive up  to an
energy scale $M'_F \le  M_F$.  Approximating the  sum over the  $n$-KK
states in Eq.\ (\ref{YBtot}) by an integral, we obtain
\begin{equation}
  \label{BAU}
Y_{\Delta B}\ \approx\ \frac{1}{3}\, e^{-T_c/T_r}\, S_\delta\,
\varepsilon_\chi\, \ln\bigg(\frac{M'_F}{T_{\rm min}}\bigg)\, ,
\end{equation}
with  $T_{\rm  min} ={\rm max}\,   (T_r,  m_{\rm min})$.   {}From Eq.\
(\ref{BAU}),  we find that  the  generated  BAU, $Y_{\Delta B}$,  does
crucially   depend on $T_r$.   Considering  resonant conditions for CP
violation, i.e.\  $\varepsilon_\chi \approx  1$,  one needs   $T_c/T_r
\stackrel{<}{{}_\sim} 20$ in order to generate a baryonic asymmetry at
the observed level, namely  $Y_{\Delta B} \approx 10^{-10}$.  Thus, if
the  critical temperature  is $T_c  \approx  100$  GeV, then  a reheat
temperature as low as 5--10 GeV would be sufficient to account for the
BAU, through the    mechanism of baryogenesis through    leptogenesis.
According to estimates in \cite{ADD1,BD},  to get a reheat temperature
as   high as  10  GeV  in  a theory   with $\delta   = 6$ large  extra
dimensions,  one must have  at least $M_F\approx  100$  TeV, while for
$\delta =4$ and 2 extra dimensions, the scale of quantum gravity $M_F$
must be larger than $10^4$ and $10^6$ TeV, respectively.  

Another difficulty of the leptogenesis models  we have been discussing
is  that the late decays  of  the low KK   neutrinos  may distort  the
abundances related to  the light elements  $^4$He, D, $^7$Li, etc.  Of
course, for sufficiently  large  values of $M_F$ and/or  $\delta$, the
lowest  KK state,  with  mass of  order $1/R \sim (M_F/M_P)^{2/\delta}
M_F$, will be heavy enough to decay just before nucleosynthesis.  This
may reintroduce  a mild  hierarchy problem in  the  parameters of  the
theory,  in  case  we  wish  to identify  $M_F$   with  the  scale  of
soft-supersymmetry   breaking  \cite{ADD}.      Therefore,  among  the
leptogenesis  models that  were discussed   in  Section 3, the  hybrid
scenario represents the  most attractive solution  to this problem, as
the lowest KK   state, with  mass $m_{\rm   min} = m-M$,   can be made
sufficiently heavy in order  to decay rapidly  enough.  Whether such a
scenario can be embedded to a more general supersymmetric theory is an
issue that we shall not address in the present work.

\section{Conclusions}

We  have studied the  scenario of baryogenesis through leptogenesis in
theories   with large  compact  dimensions.  The  formulation of these
theories requires the extension  of the notion  of the Majorana spinor
to multidimensional Minkowski spaces.  We  have reviewed this topic in
Section  2. In particular, it  was shown that genuine massive Majorana
neutrinos exist in 2, 3 and 4$\, {\rm mod}\, 8$ dimensions only.  This
limitation is   due      to the lack  of      finding Clifford-algebra
representations that satisfy the  Majorana properties in any number of
dimensions.  In   Section  3, we   have formulated minimal   models of
leptogenesis that are renormalizable if a  finite number of KK states
are considered.  Such a truncation of the  number of the KK states may
not be very unrealistic,  as the fundamental  scale of quantum gravity
$M_F$ is expected to play the role of an UV cutoff.

Initially, the  leptogenesis  models   that we  have   been discussing
furnish the field content of the theory with an  infinite series of KK
Dirac states.  Subsequently, each KK  Dirac state splits into pairs of
nearly degenerate Majorana  neutrinos.  After  compactification of the
extra  dimensions,  such a mass   splitting occurs either  at the tree
level, or, more interestingly,  at  the one-loop level  by integrating
out   the Higgs interactions   of  an  extended  Higgs  sector.   As a
consequence, each    pair  of the Majorana   neutrinos   behaves  as a
CP-violating  resonator, i.e.\ it  becomes  a strongly mixed two-level
system producing a leptonic CP  asymmetry of order unity. Depending on
the   mass  difference between two    adjacent pairs   of KK  Majorana
neutrinos,  the     tower of  CP-violating     resonators  may have  a
constructive or destructive interference.  In Section 4, we have found
that such   an interference  is  constructive,  if the   level spacing
between any  two nearby pairs of  KK Majorana neutrinos is much larger
than the mass  difference of the  Majorana neutrinos within each  pair
(cf.\ Eq.\ (\ref{constr})).

In Section 5, we have seen that the KK Majorana neutrinos mostly decay
out of thermal equilibrium in  theories with large compact dimensions. 
Based on  the  afore-mentioned CP-violating  mechanism,  the resulting
leptonic asymmetry is of order unity.  However, in theories with a low
scale  of quantum gravity \cite{ADD1,ADKM}, gravitational interactions
play an   essential role,  as  they generically   lead  to  low reheat
temperatures, much below $T_c$.   Then, the conversion  of an $L$ into
$B$ asymmetry is exponentially  suppressed, as  sphalerons are out  of
thermal equilibrium.  In   such  a cosmological framework, the   upper
bound on the  reheat temperature compatible  with  baryogenesis may be
reduced by  almost one order of  magnitude relative to $T_c$. In fact,
we can estimate that  a reheat temperature $T_r$ as  low as 5--10  GeV
would be sufficient to account for  the BAU through out-of-equilibrium
sphaleron interactions.  The latter   leads to the lower limits:  $M_F
\stackrel{>}{{}_\sim} 10^6,\ 10^4,\ 100$  TeV, for $\delta =2$, 4, and
6 large extra  dimensions,   respectively.   An important  virtue   of
leptogenesis, when compared to  the usual scenarios of baryogenesis in
low-string scale theories,  is that one does  not need  to worry about
suppressing $B$-violating  interactions which might lead to observable
proton  decays   \cite{Dvali}.  For  this   reason,   we believe  that
embedding the  minimal scenarios of leptogenesis  that we have studied
here   into  more  realistic    models  of  inflation  constitutes  an
interesting issue for future investigations.

\subsection*{Acknowledgements} 
The author thanks  Peter     Breitenlohner and Ara  Ioannisian     for
discussions on the topic  of  higher-dimensional Majorana spinors  and
related issues, and Savas Dimopoulos, Gia Dvali and Antonio Riotto for
their valuable comments on inflationary models  in theories with large
compact dimensions.

\end{document}